# A biologically Inspired Trust Model for Open Multi-Agent Systems that is Resilient to Rapid Performance Fluctuations

Zoi Lygizou [1,] * and Dimitris Kalles [2,*]

[1] Hellenic Open University, Patra, Greece; zoi.lygizou@ac.eap.gr
[2] Hellenic Open University, Patra, Greece; kalles@eap.gr
* Correspondence: zoi.lygizou@ac.eap.gr (Z.L.); kalles@eap.gr (D.K.)

**Abstract:** Trust management provides an alternative solution for securing open, dynamic, and distributed multi-agent systems, where conventional cryptographic methods prove to be impractical. However, existing trust models face challenges related to agent mobility, changing behaviors, and the cold start problem. To address these issues we introduced a biologically inspired trust model in which trustees assess their own capabilities and store trust data locally. This design improves mobility support, reduces communication overhead, resists disinformation, and preserves privacy. Despite these advantages, prior evaluations revealed limitations of our model in adapting to provider population changes and continuous performance fluctuations. This study proposes a novel algorithm, incorporating a self-classification mechanism for providers to detect performance drops potentially harmful for the service consumers. Simulation results demonstrate that the new algorithm outperforms its original version and FIRE, a well-known trust and reputation model, particularly in handling dynamic trustee behavior. While FIRE remains competitive under extreme environmental changes, the proposed algorithm demonstrates greater adaptability across various conditions. In contrast to existing trust modeling research, this study conducts a comprehensive evaluation of our model using widely recognized trust model criteria, assessing its resilience against common trust-related attacks while identifying strengths, weaknesses, and potential countermeasures. Finally, several key directions for future research are proposed.

**Keywords:** computational trust models; open multi-agent systems; biologically inspired algorithms, providers' dynamic behavior, comprehensive evaluation

## 1. Introduction

Conventional cryptographic methods, such as certificates, digital signatures, and Public Key Infrastructure (PKI) depend on a static, centralized authority for certificate verification. However, this approach may be impractical or insecure in open, dynamic and highly distributed networks [1]. Additionally, entities in such networks may have limited resources, making it difficult for them to handle the computational demands of cryptographic protocols [2]. To overcome these challenges, trust management provides an alternative approach, enabling entities to gather reliable and accurate information from their surrounding network.

Various methods for evaluating trust and reputation have been developed for real-world distributed networks that can be considered multi-agent systems (MAS), such as peer-to-peer (P2P) networks, online marketplaces, pervasive computing, Smart Grid, the Internet of Things (IoT), and many more. However, existing trust management approaches still face major challenges. In open MAS, agents frequently join and leave the system, making it difficult for most trust models to adapt [3]. Additionally, assigning trust values in the absence of evidence and detecting dynamic behavioral changes remain unresolved issues [4]. Trust and reputation models, which rely on data-driven methods, often struggle when assessing new agents with no prior interactions. This issue, known as the cold start problem, commonly arises in dynamic groups where agents may have interacted with previous neighbors but lack connections to newly introduced agents. Moreover, in dynamic environments, agents' behaviors can change rapidly, necessitating that consumer agents recognize these changes to select reliable providers. The challenge in trust assessment lies in the fact that behavioral changes occur at varying speeds and times, making conventional methods, such as forgetting old data at a fixed rate, ineffective [4].

To address these challenges, we previously introduced a novel trust model for open MAS, inspired by synaptic plasticity, the process that enables neurons in the human brain to form structured groups known as assemblies, which is why we named our model "Create Assemblies (CA)". CA's distinguishing characteristic is that, unlike traditional trust models where the trustor selects a trustee, CA allows the trustee to determine whether it has the necessary skills to complete a given task. To briefly describe the CA approach, we note that activities are initiated by a service requester, the trustor, which broadcasts a request message that includes task details such as its category and specific requirements. Upon receiving the request, potential trustees (i.e., service providers within the service requester's vicinity) store the message locally and establish a connection with the trustor if one does not already exist. Each connection is assigned a weight, a value between 0 and 1, representing the trust level or the probability of the trustee successfully completing the task. In CA model, trust update is event-driven. After the task completion, the trustor provides performance feedback, which the trustee uses to adjust the connection weight – increasing it for successful execution and decreasing it otherwise.

The CA approach offers several key advantages in open and highly dynamic MAS due to its unique design, where the trustor does not select a trustee, and all trust-related data concerning the trustee is stored locally within the trustee. First, since each trustee retains its own trust information, this data can be readily accessed and utilized across different applications or systems the agent joins. This provides a major advantage in handling mobility, an ongoing challenge in MAS [5]. By allowing a new service provider to estimate its performance for a given task based on past executions in other systems, the CA model effectively addresses the cold start problem. Second, in conventional trust models, trustors must collect and exchange extensive trust information (e.g. recommendations) before making a selection, leading to increased communication overhead. This is particularly problematic in resource-constrained networks, where communication is more expensive than computation [2]. Previous research [6] highlights that task allocation in such settings is a complex combinatorial problem, requiring multiple message exchanges. In contrast, the CA model reduces this burden by limiting communication to request messages from service requesters and feedback messages containing performance ratings, significantly minimizing communication time and overhead. Third, since agents in the CA model do not share trust information, the approach is inherently more resistant to disinformation tactics, such as false or dishonest recommendations, which are common in other trust-based approaches. However, as we will discuss later, there are still some potential vulnerabilities that need to be addressed. Finally, privacy concerns often discourage agents from sharing trust-related data in traditional models. Since CA does not rely on recommendations, agents do not have to disclose how they evaluate others' services, thereby preserving their privacy.

In our previous work [7], we proposed a CA algorithm, as a solution for managing the constant entry and exit of agents in open MAS, as well as their dynamic behaviors. Through an extensive comparison between CA and FIRE, a well-known trust model for open MAS, we found out that:

1. CA demonstrated stable performance under various environmental changes;
2. CA's main strength was its resilience to fluctuations in the consumer population (unlike FIRE, CA does not depend on consumers for Witness Reputation, making it more effective in such scenarios);
3. FIRE was more resilient to changes in the provider population (since CA relies on providers' self-assessment of their capabilities, newly introduced providers – who lack prior experience – must learn from scratch);
4. Among all environmental changes tested, the most detrimental to both models was frequent shifts in provider behavior, particularly when providers switched performance profiles, where FIRE exhibited greater resilience than CA.

Motivated by these findings, this work begins with a semi-formal analysis aimed at identifying potential modifications to the CA algorithm that could enhance its performance when dealing with dynamic trustees' profiles. Based on this analysis, we introduce an improved version of the CA algorithm with a key modification: after providing a service, a service provider re-evaluates its performance, and if it falls below a predefined threshold, it classifies itself as a bad provider. This modification ensures that each provider maintains an up-to date evaluation of its own performance, allowing for the immediate detection of performance drops that could harm consumers. Next, we conduct a series of simulation experiments to compare the performance of the updated CA algorithm against the previous version. The results demonstrate that the new version outperforms the original CA algorithm under all tested environmental conditions. In particular, when trustees change performance profiles, the improved CA algorithm surpasses both FIRE and its earlier version. We then present a comprehensive evaluation of the CA model based on thirteen widely recognized trust model criteria from the literature. While the model satisfies essential criteria such as decentralization, subjectivity, context awareness, dynamicity, availability, integrity, and transparence, further research and enhancements are identified. In this work, we have made the assumption of willing and honest agents, but we acknowledge that our solution should be able to generalize to more realistic settings, where dishonesty and unwillingness are common conditions among agents' societies. To this end, we examine CA's resilience against common trust-related attacks, highlighting its strengths and limitations while proposing potential countermeasures. To our knowledge, addressing both openness and dishonesty together is novel.

This thorough assessment not only underscores the model's strengths but also reflects a commitment to ongoing improvement, establishing it as a promising foundation for future trust management solutions in critical applications, such as computing resources, energy sharing and environmental sensing – particularly in highly dynamic and distributed environments where security and privacy are paramount. The rest of the paper is organized as follows. Section 2 reviews several trust management protocols related to our work. Section 3 provides background information concerning the CA and FIRE models, as well as the original CA algorithm. In Section 4, we conduct a semi-formal analysis of the CA algorithm's behavior, leading to the introduction of a new version that avoids unwarranted task executions. Section 5 outlines the experimental setup and the methodology for the simulation experiments, with results presented in Section 6. Section 7 offers a comprehensive evaluation of the CA model, based on widely accepted evaluation criteria, while Section 8 examines its robustness against common trust-related at-tacks. Finally, Section 9 concludes our work and suggests potential directions for future work.

## 2. Related Work

In this section, we review various trust management protocols relevant to our work. First, we identify IoT service crowdsourcing as a suitable domain for implementing trustee-oriented models, like CA. We then explore different methods proposed to address the cold start problem and agent mobility, highlighting how our approach tackles these challenges. Additionally, we review existing solutions for handling agents' dynamically changing behaviors, discuss their limitations, and explain how the CA model effectively addresses this issue. Next, we review two representative trust management protocols which, like CA, assign service providers the responsibility of assessing their ability to fulfill a requested task and deciding whether to accept or reject service requests, comparing their similarities and differences with our approach. Lastly, we emphasize that when service requesters lose control over selecting service providers, a mechanism is required to encourage honest service provision. To strengthen the CA model's robustness against trust-related attacks, we review several trust management protocols with promising mechanisms that could be integrated into our approach.

*2.1. Trust models for IoT service crowdsourcing*

The unique design of CA model makes it well-suited for IoT service crowdsourcing, where IoT devices offer services to nearby devices, known as crowdsourced IoT services. These services can include computing resources, energy sharing, and environmental sensing. For instance, in energy-sharing services, IoT devices acting as service providers can wirelessly transfer energy to nearby devices with low battery levels, which act as service consumers. Two key characteristics of IoT platforms necessitate specialized trust management frameworks, like the CA model; these are device volatility (IoT devices have short lifespan, frequently joining and leaving the network) and context-aware trust (trustworthiness in IoT environments is influenced by various factors, such as the owner's reputation, the operating system, and the device manufacturer).

The authors in [8] present a Blockchain-Based Trust Management System (BTMS) for energy trading, integrating blockchain, trust management, and MAS to enhance security and reliability. The system is structured into four layers: Blockchain Layer, which securely stores and encrypts trust values and feedback; Aggregator Layer, where coalitions of agents are managed by consensus-elected aggregators that assess trust and update blockchain records; Prosumer Layer, consisting of energy-producing and consuming agents that trade energy while exchanging trust data with aggregators; and Physical Layer, which includes distributed energy resources such as solar panels, wind turbines, and energy storage. The trust evaluation process incorporates direct and indirect assessments using a weighted average method, while multi-source feedback from aggregators is stored on the blockchain. Trust credibility is determined based on trust distortion, consistency, and reliability, ensuring accurate evaluation and identifying dishonest agents. Additionally, each agent is assigned a token, known as a Balanced Trust Incentive (BTI) unit, which is used to interact with other agents. BTI increases with honest behavior and decreases when dishonesty is detected, fostering a transparent and trustworthy energy trading environment.

The authors in [9] build upon [8] to address the unresolved problems of agent-to-agent cooperation and privacy in trust management. The BTMS leverages game theory, treating trust evaluation as a repeated game to encourage effective agent cooperation. In this framework, agents that refuse to cooperate face penalties in subsequent rounds, as others will also decline cooperation. A Tit-3-For-Tat (T3FT) strategy is em-ployed, immediately penalizing non-cooperative agents, with punishment duration depending on their level of cooperation. The defaulter agents may regain their trust after cooperating for three consecutive plays. Afterwards, their cheating behavior is forgiven. To ensure secure and verifiable trust evaluation, a Publicly Verifiable Secret Sharing (PVSS) mechanism is implemented. It distributes sensitive information, such as cryptographic keys, among three entities: dealers (blockchain miners) who allocate shares, participants who store shares, and a combiner (aggregator) who reconstructs the secret based on a defined threshold. PVSS enhances security by enabling public verification of share validity, deterring dishonest actions. Additionally, a Proof-of-Cooperation (PoC) consensus protocol is introduced within a consortium blockchain network to govern miner selection and block validation while fostering agent cooperation. However, this study focuses only on two trust- related attacks: bad-mouthing and on-off.

To assess context-dependent trust in dynamic IoT services, the authors in [10] propose a perspective-based trust management framework, which evaluates trust through three key perspectives: the Owner perspective (determined by social relationships and locality), the Device perspective (based on device reputation, which includes attributes like manufacturer and operating system), and the Service perspective (focuses on service reliability, measuring performance and its impact on trust). These attributes are processed using a machine learning based algorithm to build a trust model for crowdsourced IoT services. The authors highlight that in some IoT crowdsourcing applications – especially energy-sharing services – service reliability is a more critical factor than privacy. Additionally, social network-based trust models alone may be insufficient for evaluating trust between IoT service providers and consumers.

The CA model primarily relies on service reliability to assess the trustworthiness of service providers; thus, it can be widely used in IoT service crowdsourcing applications.

*2.2. The cold start problem*

Trust and reputation models typically perform poorly when dealing with newcomer service providers. This challenge, known as the cold-start problem, arises because service requesters have difficulty assessing the trustworthiness of newcomer service providers with whom they have no previous interactions but are now connected to. Stereotyping, first introduced in [11], operates on the idea that agents with similar observable characteristics are likely to behave similarly. This approach helps improve trust assessments for newcomers or agents with no prior interaction history. More recently, the cold-start problem has been addressed using machine learning, where trust models predict a newcomer's trustworthiness [12]. Additionally, roles within virtual organizations and social relationships among entities can serve as sources of trust information [5], particularly when direct or indirect trust data is unavailable. Another approach involves assigning an initial trust value until the agent can gather direct experience. This value may represent the average performance or be inferred from interactions with other agents [4].

The CA algorithm applies this principle to mitigate the cold-start problem. When a service provider receives a request to complete a specific task for a particular service requester for the first time, it establishes a connection and assigns an initial weight based on the average of existing weights for that task (from interactions with other service requesters). In essence, the CA algorithm allows the trustee to leverage prior experience from performing the same task for other trustors. If no prior knowledge is available, the algorithm sets the initial weight to a predetermined value, representing a baseline performance level.

*2.3. Agent mobility*

A major concern with trust and reputation systems is their lack of interoperability across applications, known as the issue of mobility of agents. When entities move between different applications, their trust and reputation data do not follow them; instead, this information remains isolated, requiring entities to rebuild their reputation in each new application. Enabling the transfer of trust and reputation data across applications remains an open challenge.

To address this issue, the authors in [5] propose a general framework to enable the exchange of trust and reputation information. This framework establishes a set of messages and a protocol that allows trust and reputation systems to request ratings, provide responses, and communicate errors. Additionally, to streamline message creation, they developed a grammar for generating queries from strings and introduced a procedure for parsing these query messages.

In [2], a hybrid trust management framework enables trust evaluation between entities using both centralized and distributed approaches, specifically addressing agent mobility. The network is divided into multiple sub-networks to improve management efficiency and scalability. Each sub-network contains a resource-rich fog node, known as the kingpin node, responsible for sharing service provider reputation scores upon request. For global reputation assessment, kingpin nodes aggregate reputation scores and forward them to the cloud. When a service provider moves to a different sub network, it can present the ID of the kingpin from its previous sub network. The new kingpin can then retrieve the service provider's reputation from the cloud using this ID.

In the CA approach, however, a service provider retains all trust-related information concerning itself in the form of stored connections. This allows the service provider to directly access its own trust data and manage mobility by calculating its trustworthiness as a global reputation.

*2.4. Dynamic changes in agents' behaviors*

In dynamic environments, agents' behavior can change rapidly, making it essential for consumer agents to detect these changes when using trust and reputation models to select reliable provider agents. Behavioral changes may stem from malicious intent, but they can also result from resource limitations, such as reduced battery power [2]. In MAS, dynamic behavior can be influenced by various factors, including seasonal variations, device malfunctions, malicious actions, network congestion, and more [4]. Some changes occur randomly and unpredictably, while others follow cyclical patterns. Additionally, group dynamics can contribute to behavioral changes, as relationships between agents evolve due to changing motivations or external influences, such as environmental changes.

An example of agents exhibiting continuously changing behavior is found in fifth-Generation (5G) networks, where forwarding entities are responsible for relaying packets along a route. These entities may drop packets, leading to a reduced packet delivery rate. Their behavior is inherently dynamic, as they can switch between malicious and legitimate states at any time [13]. Reinforcement Learning (RL) allows legitimate forwarding entities to learn about the behavior of other potential forwarders. However, malicious entities can also exploit RL to strategically drop packets while avoiding detection, thereby manipulating the learning environment of legitimate entities. To address this issue, the authors in [13] propose a hybrid trust model for secure routing in 5G networks. In this model, Q-learning enables distributed legitimate entities to evaluate the trustworthiness of neighboring forwarding entities. Additionally, using network-wide trust data and RL, a central authority can decide whether legitimate entities should continue or halt their learning process, depending on the proportion of legitimate to mali-cious entities in the network. However, the effectiveness of this model has not been experimentally validated.

To handle dynamic behavior, current trust and reputation models continue to utilize methods like sliding windows and forgetting factors [4]. A sliding window of size $n$ preserves only the $n$ most recent experiences, based on the assumption that these interactions best reflect an agent's current behavior, while older records are discarded as they become less relevant. A forgetting factor gradually decreases the influence of past interactions over time, retaining all instances but assigning greater weight to more recent experiences. In [14], the proposed trust model for pervasive computing applications is an instance of the forgetting factor approach, by progressively reducing the influence of past interactions through referral age in social trust calculations. It integrates both personal experience and reputation using Fuzzy Logic (FL) principles. The model consists of three subsystems, with two of the subsystems responsible for evaluating an entity's trust level based on individual experiences and referrals. To address the dynamic nature of trust, the calculation of social trust depends on two key factors: the referral age, and the referrer reputation. Each referral is timestamped, and its influence on social trust decreases over time. Once it surpasses a predefined expiration point, it is no longer considered. The credibility of the referrer determines how much weight their referral carries in the final trust calculation. The system employs a central authority to manage storage and access to referrals. However, this introduces a single point of failure, which can be a limitation. After determining social and individual trust, each intelligent agent uses a weighted sum method to compute the final trust value. The third FL subsystem is responsible for determining the weight assigned to social trust, using as inputs the total number of social referrals and the total number of past individual interactions.

Assessing trust in MAS is particularly challenging due to the dynamic nature of agent behavior, which can change at varying speeds and times. This variability makes traditional approaches, such as forgetting old data at a constant rate and sliding windows, ineffective [4]. To address this issue, the authors in [4] propose RaPTaR, a framework that sits on top of existing trust and reputation models to detect and adapt to behavior changes within agent groups. RaPTaR enhances trust algorithms by supplying past experience data that has been statistically evaluated to reflect an agent's current behavior. The system identifies behavioral shifts by analyzing the outcome patterns of agent groups using the Kolmogorov-Smirnov (K-S) statistical test. Additionally, it records transitions between behavior changes to recognize recurring patterns, which can be leveraged for future predictions. By exploiting repetitive behavior, RaPTaR improves trust assessment in dynamic environments. Experimental results indicate that while RaPTaR effectively detects and responds to random behavioral changes, its accuracy remains limited and can be improved.

Reputation-based trust models are widely used in distributed networks, but they encounter two major challenges: The transmitted reputation may not accurately reflect an entity's current trustworthiness, and over-reliance on reputation makes the system vulnerable to collusion, where malicious entities conspire to fabricate trust values. One possible solution is trust prediction, which estimates an entity's current trustworthiness based on past interactions, behavioral history, and other objective factors. Traditional time-based prediction methods fall into two categories. The one is complete arithmetic mean, which computes the average of all past data to predict the next trust value. The other method is Mean shift, which prioritizes recent data, ignoring long-term history. A more refined approach is exponential smoothing, which balances both methods by incorporating past data while assigning diminishing weight as time elapses. However, pure exponential smoothing struggles with highly volatile data, especially in cases where a service provider is hijacked by adversaries, causing an abrupt shift from high trust to deception. In such situations, exponential smoothing alone can lead to significant deviations in trust estimation. To address this issue, the authors in [15] propose a dynamic trust model that integrates direct and indirect trust computation with trust prediction. Their approach first applies exponential smoothing to compute a trend curve, and then utilizes a Markov chain model to adjust for fluctuations, improving prediction accuracy. This is a refined method for handling abrupt behavioral shifts, such as when an entity transitions from trustworthy to malicious.

In the proposed CA algorithm, after delivering a service, a service provider reassesses its performance. If the performance falls below a predefined threshold, the service provider categorizes itself as a bad provider. This ensures that each provider continuously maintains an updated assessment of its capabilities, enabling the prompt detection of performance declines that could negatively impact consumers.

*2.5. Assessing trust from the trustee's perspective*

Similar to the CA approach, several studies [16, 17] place the responsibility on service providers to determine whether they can complete a requested task and assess the potential benefits before deciding whether to accept or decline service requests. In [16], the author introduces ConTrust, a context-aware trust model for Social Internet of Things (SIoT), designed to help service consumers select reliable providers for task assignments. In this approach, the requester broadcasts the task, specifying its requirements, and providers evaluate their ability to fulfill the request along with the potential benefits. If a provider is willing to take on the task, it responds to the requester, who then assesses the trustworthiness of the candidates and selects a specific provider. Trustworthiness in ConTrust is measured using a weighted sum of three factors: service provider's capability, commitment, and job satisfaction feedback. The CA model shares similarities with ConTrust in that it is also context-aware, as trust evaluation considers the task type and requirements within the specific environment where the service is provided. Additionally, the consumer broadcasts a request, and each provider evaluates its own capability to complete the task. However, the CA model differs in a fundamental way: service consumers do not select providers. Instead, trust assessment is handled by the provider itself, based on consumer feedback.

A key issue with trust and reputation systems that rely solely on distributed infrastructures is that each entity only retains information about a small fraction of the entire network. As a result, knowledge about the number of available service providers is limited to local information, reducing the likelihood of selecting highly trustworthy providers [2]. This limitation does not apply to the CA approach. In CA, the request message is broadcasted to all service providers within the consumer agent's vicinity. However, unreliable providers, who are aware of their own limitations, will opt out of executing tasks they are not capable of handling. Consequently, this self-selection process increases the probability of obtaining service from highly trustworthy service providers.

When a service requester no longer has control over selecting service providers, and instead, the service providers themselves decide whether to offer a service, a mechanism is needed to incentivize honest service provision. Additionally, most trust and reputation models assume that service requesters act honestly, which is often unrealistic in agent-based systems. To address these issues, [17] proposes a blockchain-based, two-way reputation system for SIoT, incorporating a penalty mechanism for both dishonest service providers and service requesters. This approach evaluates a service provider's local trust, global trust, and reputation by considering both Social Trust and Quality of Service (QoS) factors, such as Availability, Accuracy, Cruciality, Responsiveness, and Cooperation. After receiving a service, service requesters must provide feedback by rating the service providers. However, relying on a single cumulative rating has drawbacks: i) it does not clarify the evaluation criteria used for assessing service providers, ii) dishonest service requesters can intentionally lower a service provider's rating, even if the service is good, and iii) inexperience service requesters may struggle to provide meaningful feedback. To overcome these challenges, the authors propose a "two-stage parameterized feedback" system. Tlocal (local trust) is calculated in two phases: pre-service avail and post-service avail. Comparing Tlocal and Tglobal (global trust) helps detect suspicious or dishonest service requesters, who are then classified into three categories: suspicious, temporarily banned (can only request certain services), and permanently banned (prohibited from requesting any service). For service providers, a penalty mechanism using a fee charge system imposes monetary losses on dishonest providers. Service providers are categorized into three status lists based on their reputation: white list (highest service fees and selection priority), grey list (moderate reputation, lower fees), and black list (least trustworthy, low selection probability). Each feedback update adjusts a service provider's trust and reputation value, and service providers can accept or decline service requests. The approach also allows service requesters to choose service providers based on cost preference – for example, a service requester aiming to reduce expenses may prefer a service provider from the Grey List rather than the White List. Experimental results show that this method is resilient to various trust-related attacks, including On-Off, Discriminatory, Opportunistic Service, Selective Behavior, Bad Mouthing, and Ballot Stuffing Attacks. However, the study does not address the Whitewashing Attack. The proposed mechanisms offer valuable insights that could serve as a foundation for adapting the CA approach to more realistic environments where dishonest behavior is prevalent.

*2.6. Mechanisms for promoting honest behavior*

Besides [17], several other studies [18-24] propose mechanisms aimed at promoting honest behavior among agents during their interactions. These mechanisms serve as a basis for formulating a customized approach within the CA framework to counter various trust-related attacks, which we explore further in Section 8.

A key requirement in large, dynamic, and heterogeneous networks –where node participation is constantly changing – is a mechanism for establishing a secure, authenticated channel between any two participating nodes to exchange sensitive information. The study in [18] presents a Bayesian trust model that probabilistically estimates node trust scores in an IoT environment, even in the absence of complete information. The authors propose a contract-theoretic incentive mechanism to build trust between devices in an ad-hoc manner by leveraging locally adjacent nodes. Each IoT node independently stores and advertises its absolute trust score. The proposed effort-reward model motivates selected nodes to accurately report their trust scores and actively contribute to the authentication process, with rewards aligned to their actual trust levels. Unlike traditional blockchain-based solutions, this approach does not depend on cryptographic primitives or a central authority, and the final consensus is localized rather than being a universally shared state across all nodes in the system.

To enhance content trust, the authors in [19] introduced TrustCoin, a smart trust evaluation framework for Information-Centric Networks (ICNs) in Beyond Fifth-Generation (B5G) networks, leveraging consortium blockchain technology. In this scheme, each TrustCoin user –whether a publisher/producer or a subscriber/consumer – is assigned credit quotas (i.e., coins) that reflect its reputation and serve as a measure of trust. A higher coin balance indicates greater credibility. Users must first register to obtain a legal identity and initial credit coins before they can request to publish content or report potentially malicious data. When content is published, a checking server verifies the publisher's credit balance on the blockchain to ensure it meets a predefined threshold. For content sharing, edge nodes authenticate user identities, while Deep Rein-forcement Learning (DRL) is employed to assess content credibility. TrustCoin incorporates an incentive mechanism that encourages users to proactively publish trustworthy content. The system updates the credit coins of publishers and subscribers based on content credibility determined by DRL-driven evaluations. Higher credibility results in rewards for publishers, reinforcing trustworthy behavior, while subscribers may face penalties. Conversely, if a publisher shares low-credibility content, they incur a penalty, while the subscriber is rewarded for correctly identifying and reporting it.

In IoT networks, connected nodes may exhibit reluctance to forward packets in order to conserve resources such as battery life, energy, bandwidth, and memory. To address this issue, the authors in [20] propose a trust-based approach called HBST, which fosters cooperation by forming a credible community based on honesty. Unlike conventional methods that permanently remove selfish nodes from the network, HBST instead isolates them in a separate domain, preventing interactions with honest nodes while offering then an opportunity to improve their behavior and reenter the network. The proposed model consists of two phases. In the first phase, credible nodes – those with a sufficient community reputation – are selected from the main network to form a "credible community". A node's reputation is determined by its honesty level, assessed using metrics such as interaction frequency, credibility, and community engagement. In the second phase, a social selection process appoints two leaders from this credible community based on factors such as seniority, cooperative behavior, and energy levels. The node with the highest reputation weight is designated as the Selected Community Head (SCH), while the second highest becomes the Selected Monitoring Head (SMH). These leaders play key roles in mitigating selfish behavior: the SCH is responsible for encouraging selfish nodes to participate and can impose penalties on reported offenders, while the SMH monitors node behavior and flags selfish activity to the SCH. Selfish nodes are initially isolated in a separate domain, barring communication with the rest of the network. If a selfish node repeatedly fails to improve, it may face expulsion – a strict penalty. However, unlike permanent removal, HBST allows selfish nodes to rejoin the main community by improving their honesty level beyond a predefined threshold. In cases of persistent selfish behavior, a warning message is broadcasted to neighboring communities, instructing them to cease communication with the offending node as a severe consequence.

In the Internet of Medical Things (IoMT), numerous smart health monitoring devices communicate and transmit data for analysis and real-time decision-making. Secure communication among these devices is essential to ensure timely and accurate patient data processing. However, establishing reliable communication in large-scale IoMT networks is both time-intensive and energy-demanding. To address this challenge, the authors in [23] propose BFT-IoMT, a distributed, energy-efficient, fuzzy logic-based, blockchain-enabled, and fog-assisted trust management framework. This framework employs a cluster-based trust evaluation mechanism to detect and isolate Sybil nodes. The process begins with a topology lookup module, which identifies network topology whenever an IoMT device joins or leaves the network. Next, clusters and Cluster Heads (CHs) are formed, with CHs registered on the blockchain. Each IoMT node's trust-related parameters are collected and analyzed by a trust calculator module. The computed trust scores are then stored on the blockchain. If a node's trust score falls below a predefined threshold, it is flagged as malicious (Sybil) node and isolated from the network. The final decision on whether a node is malicious or benign is then broadcasted across the IoMT system. The fog-assisted trust framework is designed to enhance network throughput while reducing latency, energy consumption, and communication overhead. The use of fuzzy logic – which efficiently handles ambiguous and uncertain data, common in healthcare applications – improves the computational power and effectiveness of the decentralized trust management system. However, the proposed protocol assumes that CHs are inherently trusted nodes, which is an impractical assumption in most IoT environments.

In [24], the authors aim to develop a fully distributed and scalable trust management protocol, enabling IoT devices to assess the trustworthiness of any service provider on the Internet without relying on pre-trusted entities. They introduce a decentralized trust management protocol for the Internet of Everything (IoE), leveraging blockchain technology and the fog computing paradigm. In this system, powerful fog nodes maintain the blockchain, relieving lightweight IoT devices of the burden of trust data storage and intensive computations. This approach optimizes resource usage, bandwidth efficiency, and scalability. By utilizing blockchain, the protocol provides a global perspective on the trustworthiness of each service provider within the network. A key feature of the proposed solution is its fine-grained trust evaluation mechanism – IoT devices receive recommendations about service providers not only based on the requested service but also according to a set of specific requirements that the providers can fulfill. Additionally, blockchain technology enhances the protocol's adaptability to high-mobility scenarios. Through experiments, the authors demonstrate the resilience and robustness of their approach against various malicious attacks, including self-promotion, bad-mouthing, ballot-stuffing, opportunistic service, and on-off attacks. They further validate their results through a theoretical analysis of trust value convergence under different attack scenarios. To mitigate the impact of cooperative attacks, the authors propose an online countermeasure algorithm that analyzes the recommendation history recorded on the blockchain. This real-time algorithm is executed whenever fog nodes compute trust recommendations. Each fog node aggregates all recommendations for a particular IoT service provider and evaluates the minimum and maximum recommendation values. If the difference exceeds a predefined threshold, the provider is flagged as potentially engaging in a cooperative attack, and its recommendations are disregarded. The proposed protocol takes advantage of blockchain technology's strengths but also inherits its drawbacks. One major limitation is high resource consumption, as miners require substantial computing power to achieve consensus.

In [21], a fuzzy logic-based model is introduced to filter dishonest recommendations in the SIoT. The model evaluates recommendations based on factors such as recommendation values, the sender's location coordinates, the time of submission, and social relationships. The proposed approach detects Sybil attacks by applying fuzzy classification to received recommendations, considering their Cosine Distance and temporal proximity. The underlying assumption is that recommendations that are similar in content, closely timed, and sent from nearby locations are likely generated by the same attacker conducting a Sybil attack. Once Sybil recommendations are identified, the remaining recommendations are classified based on the existing social relationships between the senders and the recommended object. These relationships include Ownership Object Relationship (OOR), Co-Location Relationship (C-LOR), Co-Work Relationship (C-WOR), and Social Object Relationship (SOR). To further assess recommendation reliability, the model evaluates Internal Similarity (IS), which measures how closely each recommendation aligns with the median value of trusted witness objects within the same community. Simultaneously, it calculates the Degree of Social Relationship (DSR), which quantifies the strength of connections between the senders and the recommended object. DSR is determined by the number of past transactions between objects, with greater interaction frequency indicating higher trustworthiness. Additionally, recommendations from specific communities, such as OOR, carry more weight in the evaluation. The IS and DSR metrics play a crucial role in detecting good-mouthing and bad-mouthing attacks. If both IS and DSR values are very low, the recommendation is deemed unreliable, sug-gesting that the sender may be attempting to manipulate trust through either good-mouthing or bad-mouthing attack.

In Vehicular Edge Computing (VEC), edge servers may request data from Autonomous Vehicles (AVs) to support intelligent applications such as Intelligent Transportation Systems (ITS). However, economic concerns, including power consumption, create challenges in integrating data sharing within the dynamic VEC network. A promising alternative is shifting form data sharing to data trading, which incentivizes AVs to exchange their data for rewards. However, integrating data trading into edge servers introduces additional concerns related to security, privacy, and trust. To address these challenges, the authors in [22], propose a novel reputation management system for data trading that utilizes DRL. Their approach, called Dynamic Selection of Trusted Sellers using Deep Deterministic Policy Gradient (DSTSDPG), dynamically adjusts a credibility score threshold to identify the most reliable data sellers among the available AVs. The proposed system follows a hierarchical network architecture consisting of three levels: Vehicle level (Autonomous Vehicles), Edge level (Roadside Units (RSUs) or edge servers), and Cloud level (a central cloud server). Within each cluster, a designated VEC server retrieves credibility values from the cloud and uses multiple parameters to calculate an AV's credibility score using the DRL-based approach. Based on this score, the edge server selects the most trustworthy AV for data transactions. The computed scores are uploaded to the cloud and can be shared with other edge servers when an AV moves to a different cluster addressing the issue of mobility. However, since the method relies on a centralized architecture involving both cloud and edge servers, it inherits typical limitations such as the risk of a single point of failure. For credibility score calculation, the proposed approach considers three key factors: historical reputation (past interactions influence trustworthiness), familiarity (higher values indicate an edge server has more prior knowledge of an AV), and freshness (recent interactions are weighted more heavily). The proposed trust and reputation model is built on subjective logic, and the credibility score is derived from a weighted sum of familiarity and freshness. However, the authors do not clarify how the weights are determined.

Existing trustworthiness models can only detect a subset of known attacks, but none can defend against all types [21]. This is because attack patterns are highly diverse, with malicious nodes strategically exploiting weaknesses in trust algorithms to evade detection. To address this challenge the authors in [21] examine all trust-related attacks documented in the literature that can impact IoT systems. They propose a decentralized trust management model that leverages a Machine Learning algorithm and introduces three novel parameters: goodness, usefulness, and perseverance scores. These scores enable the model to continuously learn, adapt, and effectively identify and counteract a wide range of malicious attacks. Therefore, developing algorithms capable of detecting diverse malicious activity patterns is crucial [25]. To this end, in Section 8, we further discuss how various mechanisms can be integrated into the CA approach to enhance its resilience against existing trust-related attacks.

## 3. Background

In this section, we begin by outlining the key features of FIRE and CA, the two trust models evaluated in our simulation experiments, explaining the rationale behind selecting FIRE as a reference model for comparison, followed by an overview of the previous version of the CA algorithm, which incorporates a mechanism for handling dynamic trustee profiles.

*3.1. FIRE*

We selected FIRE [26] as a reference model for comparison with CA because it is a well-established trust and reputation model for open MAS that, like CA, adopts a decentralized approach. Additionally, FIRE represents the traditional trust management approach, where trustors select trustees, providing a contrast to the CA model, where trustees are not chosen by trustors. It consists of four key modules:

- Interaction Trust (IT): Evaluates an agent's trustworthiness based on its past interactions with the evaluator;
- Witness Reputation (WR): Assesses the target's agent's trustworthiness using feedback from witnesses – other agents that have previously interacted with it;
- Role-based Trust (RT): Determines trustworthiness based on role-based relationships with the target agent, incorporating domain knowledge such as norms and regulations.
- Certified Reputation (CR): Relies on third-party references stored by the target agent, which can be accessed on demand to assess trustworthiness.

IT is the most reliable trust information source, as it directly reflects the evaluator's satisfaction. However, if the evaluator has no prior interactions with the target agent, FIRE cannot utilize the IT module and must rely on the other three modules, primarily WR. However, when a large number of witnesses leave the system, WR becomes ineffective, forcing FIRE to depend mainly on the CR module for trust assessments. Yet, CR is not highly reliable, as trustees may selectively store only favorable third-party references, leading to an overestimation of their performance.

*3.2. CA model*

CA is a biologically inspired computational trust model. It is inspired by biology concepts, particularly synaptic plasticity in the human brain and the formation of assemblies of neurons. In our previous work [27], we provided a detailed explanation of synaptic plasticity and how it is applied in our model.

A MAS is used to represent a social environment where agents communicate and collaborate to execute tasks. In [27], we formally defined the key concepts necessary for describing our model, but here, we summarize the most essential ones. A Trustor is an agent that defines a task and broadcasts a request message containing all relevant details, including: i) the task category (type of work to be performed), and ii) a set of requirements, as specified by the Trustor.

In CA approach, the Trustee, rather than the Trustor, decides whether to engage in an interaction and perform a given task. When a Trustee receives a request message, it establishes and maintains a connection weight $w \in [0,1]$, representing the strength of this connection. This weight reflects the probability of successful task completion and is updated by the Trustee, based on performance feedback provided by the Trustor. If the Trustee successfully completes the task, the weight is increased according to:

$$w = Min\big(1, w + \alpha(1 - w)\big). \tag{1}$$

If the Trustee fails, the weight decreases as follows:

$$w = Max\big(0, w - \beta(1 - w)\big), \tag{2}$$

where α and β are positive parameters controlling the rate of increase and decrease, respectively.

The Trustee decides whether to accept a task request by comparing the connection weight with a predefined $Threshold \in [0,1]$. If the weight meets or exceeds this threshold ($w \geq Threshold$), the Trustee proceeds with the task execution.

*3.3. The CA algorithm to handle dynamic trustee profiles*

In this subsection, we present the original version of the CA algorithm, as introduced in [7], including a brief description and the pseudocode, for the sake of completeness and to facilitate comparison with the updated version, presented in Section 4.2 of this work.

When a trustor identifies a new task to be executed by a trustee, it broadcasts a request message containing relevant task details (lines 2-3). Upon receiving this request, each potential trustee stores it in a list and establishes a new connection with the requesting trustor if one does not already exist (lines 4-11). At each timestep, each trustee reviews the tasks stored in that list and attempts to execute the task with the highest connection weight, provided that the task remains available and the weight does not exceed a predefined Threshold (lines 12-20). If the task execution is successful, the trustee increases the connection weight; otherwise, it decreases it (lines 21-25).

Lines 6-11 address the cold start problem, which arises when a trustee lacks prior experience with a specific task and thus cannot assess its own capability to complete it. In the version shown in Algorithm 1, the trustee leverages prior knowledge gained from performing the same task for other trustors. If no prior experience exists (i.e., the trustee has never performed the task for any trustor), the connection weight is initialized to a default value of 0.5 (line 9), as was done in the original algorithm proposed in [27].

This CA algorithm also accounts for dynamic trustee profiles as follows (lines 26-29). If a connection's weight falls below the Threshold, the trustee interprets this as an indication of its incapability to complete the task successfully and stops attempting it to save resources. However, if the trustee performs well on an easier task, it may infer that it has likely learned how to execute more complex tasks within the same category. In such cases, it increases the connection weight of those previously failed tasks to the Threshold value, allowing itself an opportunity to attempt them again in the future.

---

ALGORITHM 1: CA v2, for agent $i$

---

1: **while** True **do**

    # broadcast a request message when a new task is perceived

2:   **when** perceived a new task $= (c, r)$

3:     broadcast message $m = (request, i, task)$

    #receive/store a request message and initialize a new connection

4:   **when** received a new message $m = (request, j, task)$

5:     add $m$ in list $M$

6:     **if** $\nexists$ connection $co = (i, j, \_, task)$ **then**

7:       **if** $\exists$ connection $co' = (i, \sim j, \_, task)$ **then**

8:         create $co = (i, j, avg, task)$, s.t. $avg = \frac{\sum_{\forall (i,\sim j,w,task) \in \{(i,\sim j,\_,task)\}} w}{|\{(i,\sim j,\_,task)\}|}$

9:       **else** create $co = (i, j, 0.5, task)$

10:     **end if**

11:   **end if**

    #select a task to perform

12:     select message $m = (request, j, task)$ in $M$, s.t.

  ∃ co = (i, j, w, task) ∈ {(i, k, w$'$, task)}, ∀w$'$ : w ≥ w$'$

13:   if state_visibility(task) = false or state_done(task) = false then

14:     if canAccessAndUndertake(task) = true then

15:       if w ≥ Threshold then

16:         (result, performance) ← performTask(task)

17:       end if

18:     end if

19:   end if

20:   delete m from M

  #update connections' weights

21:   if result = success then

22:     strengthen co = (i, j, w, task) by using equation  (1)

23:   else

24:     weaken co = (i, j, w, task)  by using equation  (2)

25:   end if

26:   ∀ co$'$ = (i, j, w$'$, task$'$), s.t. w$'$ < Threshold, task$'$ = (c, r$'$), r$'$ > r

27:     if performance ≥ minSuccessfulPerformance(task$'$) then

28:       w' ← Threshold

29:     end if

30: end while

## 4. Enhancing performance by avoiding unwarranted task executions

In this section, we begin with a semi-formal analysis of the CA algorithm to identify potential modifications that could enhance its performance. Following this, we introduce an improved CA algorithm designed to detect and prevent unwarranted task executions.

*4.1. A semi-formal analysis of the CA algorithm to identify and avoid unwarranted task executions*

To identify potential improvements to the performance of the CA algorithm, we conducted the following semi-formal analysis. While this analysis incorporates formal elements such as assumptions, definitions, propositions, and a proof by induction, it does not constitute a strictly formal analysis in the traditional mathematical or computer science sense. Instead, it represents a blend of formal reasoning and empirical evaluation for the following reasons:

1. Although induction is used to support Proposition 1, many conclusions rely more on empirical observation than on rigorous formal logic;
2. The conclusions are drawn from specific experimental conditions (e.g., Threshold = 0.5, a=β=0.1), whereas a purely formal analysis would typically strive to derive results that hold regardless of particular parameter values;
3. Several conclusions are derived from example scenarios rather than universally valid logical proofs.

In our testbed, each consumer requesting the service at a certain performance level, waits for a period equal to WT for the service to be provided by a provider in its operational range (nearby provider). If the requested $task = (service\_ID, performance\_level)$, where $performance\_level \in \{PERFECT, GOOD, OK, BAD, WORST\}$ is not performed, either because there are no nearby-providers, or none of the nearby providers are willing to per-form the task, then the consumer requests the service at the next lower performance level, assuming that any consumer can manage with a lower-quality service.

**Assumption 1**. *WT is large enough so that if there is a capable provider in the operational range of the consumer, then the task will be successfully executed.*

**Assumption 2**. *The reason why a nearby provider may not be willing to perform a task is only because the weight of the relevant connection is less than the Threshold vale. We assume that there are no other reasons, i.e., the providers are not selfish or malicious. In other words, we assume that the providers are always honest and comply with the implemented protocols having no intention to harm the trustors.*

Since every rational consumer aims to maximize its profit, the sequence of tasks that it may request until a provider is found to provide the service is: $task_1$, $task_2$, $task_3$, $task_4$, and $task_5$. In Table 1 the requirements of each task are specified.

**Table 1.**

| Tasks | Explanation of the Requirement |
|---|---|
| task1 = (service_ID, performance_level = PERFECT) | Performance must be equal to 10 |
| task2 = (service_ID, performance_level = GOOD) | Performance must be greater than or equal to 5 |
| task3 = (service_ID, performance_level = OK) | Performance must be greater than or equal to 0 |
| task4 = (service_ID, performance_level = BAD) | Performance must be greater than or equal to -5 |
| task5 = (service_ID, performance_level = WORST) | Performance must be greater than or equal to -10 |

**Definition 1.** *Given that $C_j$ is a consumer requesting the successful execution of $task_i$, $i \in \{1,\cdots,5\}$, $P_k$ is a provider in the operational range of $C_j$ receiving a request from $C_j$ for the execution of $task_i$, and $P_k$ does not have yet a connection for $C_j$ and $task_i$, we define that $P_k$ has learned its incapability to perform $task_i$ (meaning that it cannot perform $task_i$ always successfully) when it initializes the weight of a new connection for $P_k$ and $task_i$ to a value smaller than the Threshold value, which will result in not executing $task_i$ for $C_j$.*

Since only bad and intermittent providers have negative (causing damage to the consumers) performances, we analyze how bad providers learn their own capabilities in each of the five tasks, aiming to identify ways to reduce task executions with negative performances and thus improve the performance of the CA algorithm.

We consider a system of four consumers: $C_1$, $C_2$, $C_3$, $C_4$ and only one provider: the bad provider $BP_1$. Each consumer has in its operational range the provider $BP_1$. The provider has no knowledge of its capabilities in performing tasks, meaning that it has not formed any connections yet.

Phase A: Provider $BP_1$ learns its incapability in performing $task_1$.

Assume that $C_1$ requests the execution of $task_1$ broadcasting the message $m_1 = (request, C_1, task_1)$. According to the CA algorithm, when $BP_1$ receives message $m_1$ it will create the connection $co_1 = (BP_1, C_1, 0.5, task_1)$, initializing its weight to the value 0.5. Since the condition $w \geq Threshold$ is satisfied, $BP_1$ will perform $task_1$, but it will fail because its performance range is in $[-10,0]$ and $task_1$ requires a performance equal to 10. After it fails, $BP_1$ will decrease the weight of $co_1$ to the value $w' = 0.45$, by using equation $w' = w - \beta(1 - w)$, where $\beta = 0.1$ in our experiments.

Now, let $C_2$ require the execution of $task_1$ by sending the message $m_2 = (request, C_2, task_1)$. When $BP_1$ receives the message $m_2$ and because it already has the connection $co_1$, it will create the connection $co_2 = (BP_1, C_2, 0.45, task_1)$, initializing its weight to the average of the weights of the connections it has with other consumers for $task_1$. In this case, $average = \frac{w_{co_1}}{1} = 0.45$, where $w_{co_1}$ denotes the weight of connection $co_1$. Because the weight of $co_2$ is less than the Threshold value, $BP_1$ will decide not to perform $task_1$ for consumer $C_2$.

**Assumption 3.** *For simplicity, we assume that $BP_1$ will not change its ability to provide $task_1$, so that the weights of all connections will remain constant over time.*

Now, we can prove by induction the following proposition.

**Proposition 1.** *Given assumption 3, every new connection that $BP_1$ will create for $task_1$ will be initialized to the average of the weights of the connections it has already created for $task_1$, which will remain constant and equal to 0.45.*

The proof is provided in the Appendix. By proposition 1 and definition 1, it follows that $BP_1$, in a single trial, has learned that it cannot successfully perform $task_1$. We can generalize for every bad provider to the following conclusion.

**Conclusion 1.** *Given our experiments conditions (i.e. Threshold = 0.5, α=β=0.1), every bad provider needs only one trial to learn that it cannot successfully perform task1.*

Phase B: Provider $BP_1$ learns its incapability in performing $task_2$.

Following the same analysis as in Phase A, we can conclude as follows.

**Conclusion 2.** *Given our experiments conditions (i.e. Threshold = 0.5, α = β = 0.1), every bad provider needs only one trial to learn that it cannot successfully perform task2.*

Phase C: Provider $BP_1$ learns its incapability in performing $task_3$.

Assume that $C_1$ requests the execution of $task_3$ broadcasting the message $m_1 = (request, C_1, task_3)$. When $BP_1$ receives message $m_1$ it will create the connection $co_1 = (BP_1, C_1, 0.5, task_3)$, initializing its weight to the value 0.5. Since the condition $w \geq Threshold$ is satisfied, $BP_1$ will perform $task_3$. Due to its performance range in $[-10,0]$ and task's requirement that performance must be bigger or equal to zero in order to be successful, $BP_1$ has a small probability of having a performance equal to zero and thus a successful execution of $task_3$.

So, consider a scenario where the performance of $BP_1$ on $task_3$, on its first trial is 0. After it succeeds it will increase the weight of $co_1$ to the value $w' = 0.55$ by using equation $w' = w + a \cdot (1 - w)$, where $\alpha = 0.1$ in our experiments. Now, let $C_2$ request the execution of $task_3$ broadcasting the message $m_2 = (request, C_2, task_3)$. When $BP_1$ receives message $m_2$ and because it already has the connection $co_1$, it will calculate the average weight of existing connections as $average = \frac{w_{co_1}}{1} = 0.55$. Then, it will create the connection $co_2 = (BP_1, C_2, 0.55, task_3)$ initializing its weight to the average just calculated. Since the condition $w \geq Threshold$ is satisfied, $BP_1$ will execute $task_3$ for $C_2$. Suppose that $BP_1$ fails this time and decreases the weight of $co_2$ to the value $w' = 0.505$. The scenario continues with two consecutive failed executions of $task_3$ for consumers $C_3$ and $C_4$. In Table 2 we can see the average weights of the connections after each task execution.

**Table 2.**

| Consumer | Result | Final weight of the new connection | Average weight |
|---|---|---|---|
| $C_1$ | success | $w_{co_1} = 0.55$ | 0.55 |
| $C_2$ | failure | $w_{co_2} = 0.505$ | 0.5275 |
| $C_3$ | failure | $w_{co_3} = 0.48025$ | 0.51175 |
| $C_4$ | failure | $w_{co_4} = 0.462925$ | 0.499544 |

In the scenario above, $BP_1$ needed three consecutive failed executions after a successful first execution of $task_3$ to learn (according to definition 1) that it is not capable of performing this task always in a successful way, because the average weight of its connections for $task_3$ is now less that the Threshold value. This leads us to the following more general conclusion.

**Conclusion 3.** *A successful execution of $task_3$ on the first trial of a bad provider requires a number of consecutive failed executions to learn that it cannot always execute this task successfully.*

Phase D: Provider $BP_1$ learns its incapability in performing $task_4$.

Since $BP_1$ has a performance range in $[-10,0]$ and $task_4$ requires a $performance \geq -5$ to be successful, $BP_1$ has a good probability to be successful in its first trial of $task_4$. If we repeat the analysis of phase C we will be led to the following conclusion.

**Conclusion 4.** *A successful execution of $task_4$ on the first trial of a bad provider requires a number of consecutive failed executions to learn that it cannot always execute this task successfully.*

Phase E: Provider $BP_1$ learns its capability in performing $task_5$.

Since $BP_1$ has a performance range in $[-10,0]$ and $task_5$ is successfully executed if $performance \geq -10$, $BP_1$ will always execute this task successfully.

**Conclusion 5.** *A bad provider will always execute $task_5$.*

Ideally, we would prefer that a bad provider not provide the service at all, because its poor performance harms the consumer, but providing the service at least once is required to assess the provider's capabilities. However, the previous analysis demonstrates that the CA algorithm allows the bad provider to provide the service multiple times. Despite its negative performance on $task_1$, $BP_1$ performed poorly with a negative performance in phase B as well. A more intelligent agent could consider its negative performance from the first time and decide not to execute $task_2$. Furthermore, in both phase C and phase D a successful first execution of the task requires a series of consecutive failed executions which we would like to avoid.

To this end, in the following section, we propose an improved CA algorithm designed to detect bad providers early on and prevent them from damaging the consumers with their negative performances.

*4.2. The proposed CA algorithm for the early detection of bad providers*

When created, each provider will initially believe that it is not bad (provider.bad = false), because it has not provided the service yet. Each time it provides the service, it will re-evaluate its performance. If it is less than or equal to zero, it will consider itself a bad provider (provider.bad = true). Otherwise, it will not consider itself as a bad provider (provider.bad = false). In this way, each provider maintains a current assessment of whether it is a bad provider or not, based on its last performance, which enables it to immediately detect rapid changes of its performance that may be harmful to the consumers as follows.

In lines 4-14 of the proposed algorithm, the provider uses its assessment of whether it is a bad provider or not to initialize the weight of a new connection. If it believes it is bad and the task requirement is a performance level PERFECT, GOOD, or OK (line 7), the provider initializes the weight of the connection to 0.45 (line 8). Otherwise, it follows the initialization procedure defined by the previous version of the CA algorithm (lines 9-11), as described in section 3.3.

In lines 15-19, the provider uses its assessment of whether it is bad or not to modify the weight of an existing connection. When a new request message is received from a consumer j to perform a given task (line 4), and the relevant connection exists (line 15), the provider modifies the weight of the connection to 0.45 (line 17) if it (the provider) considers itself bad and the requested task requires performance >=0 (line 16).

---

ALGORITHM 2: CA v3, for agent i

---

1: while True do
    # broadcast a request message when a new task is perceived
2:   when perceived a new task = (c, r)
3:     broadcast message m = (request, i, task)
    #receive/store a request message and initialize a new connection
4:   when received a new message m = (request, j, task)
5:     add m in list M
6:     if $\nexists$ connection co = (i, j, _, task) then
7:       if i.bad = true and task.performance_level ∈ {PERFECT, GOOD, OK} then
8:         create co = (i, j, 0.45, task)
9:       else if $\exists$ connection co' = (i, ~j, _, task) then
10:         create co = (i, j, avg, task), s.t. $avg = \frac{\sum_{\forall (i,\sim j,w,task) \in \{(i,\sim j,\_,task)\}} w}{|\{(i,\sim j,\_,task)\}|}$
11:       else create co = (i, j, 0.5, task)
12:       end if
13:     end if
14:   end if
    #modify the weight of existing connection
15:     if $\exists$ connection co = (i, j, _, task) then
16:       if i.bad = true and task.performance_level ∈ {PERFECT, GOOD, OK} then
17:         modify co = (i, j, 0.45, task)
18:       end if
19:     end if
    #select a task to perform
20:     select message m = (request, j, task) in M, s.t. $\exists$ co = (i, j, w, task) ∈ {(i, k, w', task)}, $\forall w': w \geq w'$
21:   if state_visibility(task) = false or state_done(task) = false then
22:     if canAccessAndUndertake(task) = true then
23:       if w ≥ Threshold then
24:         (result, performance) ← performTask(task)
25:       end if
26:     end if
27: end if
28: delete m from M
    #update connections' weights
29: if result = success then
30:   strengthen co = (i, j, w, task) by using equation (1)
31: else
32:   weaken co = (i, j, w, task) by using equation (2)
33: end if
34: $\forall$ co' = (i, j, w', task'), s.t. w' < Threshold, task' = (c, r'), r' > r

```
35:     if performance ≥ minSuccessfulPerformance(task') then
36:        w' ← Threshold
37:     end if
38: end while
```

## 5. Experimental setup and methodology

*5.1. The testbed*

To test the performance of the revised algorithm we have performed an extensive simulation-based experimentation, on a testbed based on the one described in [26].

The environment of the testbed consists of agents that either provide services (referred to as providers or trustees) or use these services (referred to as service requesters, consumers or trustors). For simplicity, we assume all providers offer the same service, i.e. there exists only one type of task. The agents are randomly distributed within a spherical world with a radius of 1.0. The agent's radius of operation ($r_0$) represents its capacity to interact with others (e.g. available bandwidth), and it is uniform across all agents, set to half the radius of the spherical world. Each agent has acquaintances, which are other agents located within its operational radius.

Provider performance varies and determines the utility gain (UG) for consumers during interactions. There are four types of providers: good, ordinary, intermittent, and bad, as defined in [26]. Except for intermittent providers, each type has a mean performance level $\mu_p$, with actual performance following a normal distribution around this mean. Table 3 shows the values of $\mu_p$ and the associated standard deviation $\sigma_p$ for each provider type. Intermittent providers perform randomly within the range $[PL_{BAD}, PL_{GOOD}]$. The radius of operation of a provider also represents the range within which it can offer services without a loss of quality. If a consumer is outside this range, the service quality deteriorates linearly based on the distance, but the final performance remains within $[-10, +10]$ and corresponds to the utility the consumer gains from the interaction.

**Table 3.** Profiles of provider agents (performance constants defined in Table 4).

| Profile | Performance range | σp |
|---|---|---|
| Good | [PL_GOOD, PL_PERFECT] | 1.0 |
| Ordinary | [PL_OK, PL_GOOD] | 2.0 |
| Bad | [PL_WORST, PL_OK] | 2.0 |

**Table 4.** Performance level constants.

| Performance level | Utility gained |
|---|---|
| PL_PERFECT | 10 |
| PL_GOOD | 5 |
| PL_OK | 0 |
| PL_BAD | -5 |
| PL_WORST | -10 |

Simulations in the testbed are conducted in rounds. As in real life, consumers do not require services in every round. When a consumer is created, its probability of requiring service (*activity level* $\alpha$) is selected randomly. There are no limitations on the number of agents that can participate in a round. If a consumer needs a service in a round, the request is always made within that round. The round number marks the time for any event.

Consumers fall into one of three categories: a) those using FIRE, b) those using the old version of the CA algorithm, or c) those using the new version of the CA algorithm. If a consumer requires a service in a round, it locates all nearby providers. FIRE consumers select a provider following the four-step process outlined in [26]. After choosing a provider, they use the service, gain utility, and rate the service based on the UG they received. This rating is recorded for future trust assessments. The provider is also informed of the rating and may keep it for future interactions.

CA consumers do not choose a provider. Instead, they send a request message to all nearby providers specifying the required service quality. Table 4 lists five performance levels that define the possible service qualities. CA consumers first request service at the highest quality (PERFECT). After a predetermined waiting time (WT), any CA consumer still unserved sends a new request for a lower performance level service (GOOD). This process continues until the lowest service level is reached or all consumers are served. When a provider receives a request, it stores it locally and applies the CA algorithm (CA_OLD or CA_NEW, depending on the consumer group it belongs to). WT is a parameter that defines the maximum time allowed for all requested services in a round to be provided.

We assume that any consumer can manage with a lower-quality service. This assumption does not raise an issue of unfair comparison between FIRE and CA, since it also applies to consumers using FIRE. For instance, they may end up selecting a provider – potentially the only available option – whose service ultimately delivers the lowest performance level (WORST).

Agents can enter and exit the open MAS at any time, which is simulated by replacing a number of randomly selected agents with new ones. The number of agents added or removed after each round varies, but must remain within certain percentage limits of the total population. The parameters $p_{CPC}$ and $p_{PPC}$ define these population change limits for consumers and providers, respectively. The characteristics of new agents are randomly determined, but the proportions of provider types and consumer groups are maintained.

When an agent changes location, it affects both its own situation and its interactions with others. The location is specified using polar coordinates $(r, \varphi, \theta)$, and the agent's position is updated by adding random angular changes $\Delta\varphi$ and $\Delta\theta$ to $\varphi$ and $\theta$. $\Delta\varphi$ and $\Delta\theta$ are chosen randomly from the range $[-\Delta\varphi, +\Delta\varphi]$. Consumers and providers change their locations with probabilities $p_{CLC}$ and $p_{PLC}$, respectively.

A provider's performance μ can also change by a random amount Δμ within the range $[-M, +M]$ with a probability of $p_{\mu C}$ in each round. Additionally, with a probability of $p_{ProfileSwitch}$, a provider may switch to a different profile after each round.

*5.2. Experimental methodology*

In our experiments, we compare the performance of the following three consumer groups:

- FIRE: consumer agents use FIRE algorithm;
- CA_OLD: consumers use the previous version of the CA algorithm;
- CA_NEW: consumers use the new version of the CA algorithm.

To ensure accuracy and minimize random noise, we conduct multiple independent simulation runs for each consumer group. The exact Number of Simulation Independent Runs (NSIR) varies per experiment to achieve statistically significant results. The exact NSIR values are displayed in the graphs illustrating the experimental results.

The effectiveness of each algorithm in identifying trustworthy provider agents is measured by the utility gain (UG) achieved by consumer agents during simulations. Throughout each simulation run, the testbed records the UG for each consumer interaction, along with the algorithm used (FIRE, CA_OLD, or CA_NEW).

After completing all simulation runs, we calculate the average UG for each interaction per consumer group. We then apply a two-sample t-test for means comparison [28] with a 95% confidence level to compare the average UG between:

1. CA_OLD and CA_NEW;
2. FIRE and CA_NEW.

Each experiment's results are displayed using two two-axis graphs: one comparing CA_OLD and CA_NEW, and another comparing FIRE and CA_NEW. In each graph, the left y-axis represents the UG means for consumer groups per interaction, while the right y-axis displays the performance rankings produced by the UG means comparison using the t-test. The ranking is denoted with the prefix "R" (e.g., R.CA), where a higher rank indicates superior performance. If two groups share the same rank, their performance difference is statistically insignificant. For instance, in figure 1a, at the 17th interaction (x-axis), consumer agents in the CA_NEW group achieve an average UG of 6.15 (left y-axis), and according to the t-test ranking, the CA_NEW group holds a rank of 2 (right y-axis).

All experiments use a "typical" provider population, as defined in [26], consisting of 50% beneficial providers (yielding positive UG) and 50% harmful providers (yielding negative UG, including intermittent providers).

To maintain consistency, we use the same experimental values as in [26], detailed in Table 5. Additionally, the default parameters for FIRE and CA are presented in Table 6 and Table 7, respectively.

**Table 5.** Experimental variables.

| Simulation variable | Symbol | Value |
|---|---|---|
| Number of simulation rounds | $N$ | 500 |
| Total number of provider agents | $N_P$ | 100 |
| - Good providers | $N_{GP}$ | 10 |
| - Ordinary providers | $N_{PO}$ | 40 |
| - Intermittent providers | $N_{PI}$ | 5 |
| - Bad providers | $N_{PB}$ | 45 |
| Total number of consumer agents | $N_C$ | 500 |
| Range of consumer activity level | A | [0.25, 1.00] |
| Waiting Time | $WT$ | 1000 msec |

**Table 6.** FIRE's default parameters.

| Parameters | Symbol | Value |
|---|---|---|
| Local rating history size | $H$ | 10 |
| IT recency scaling factor | $\lambda$ | -(5/ln(0.5)) |
| Branching factor | $n_{BF}$ | 2 |
| Referral length threshold | $n_{RL}$ | 5 |
| Component coefficients | | |
| - Interaction trust | $W_I$ | 2.0 |
| - Role-based trust | $W_R$ | 2.0 |
| - Witness reputation | $W_W$ | 1.0 |
| - Certified reputation | $W_C$ | 0.5 |

| Reliability function parameters | | |
|---|---|---|
| - Interaction trust | $\gamma_I$ | -ln(0.5) |
| - Role-based trust | $\gamma_R$ | -ln(0.5) |
| - Witness reputation | $\gamma_W$ | -ln(0.5) |
| - Certified reputation | $\gamma_C$ | -ln(0.5) |

**Table 7.** CA's default parameters.

| Parameters | Symbol | Value |
|---|---|---|
| Threshold | $Threshold$ | 0.5 |
| Positive factor controlling the rate of the increase in strengthening of a connection | $\alpha$ | 0.1 |
| Positive factor controlling the rate of the decrease in weakening of a connection | $\beta$ | 0.1 |

## 6. Simulation Results

This section presents the results of fourteen experiments evaluating the performance of the updated CA algorithm (CA_NEW) in comparison to both CA_OLD and FIRE. Each subsection examines different environmental conditions. The first two experiments (Section 6.1) concern scenarios where service provider performance fluctuates over time. Experiment 3 (Section 6.2) is conducted in a static environment. Experiments 4-8 (Section 6.3) test the impact of a gradually changing provider population increasing up to 30%. Experiments 9-11 (Section 6.4) examine the effects of a consumer population change up to 10%. Experiments 12 and 13 (Section 6.5) explore changes in the locations of consumers and providers. Experiment 14 (Section 6.6) evaluates performance where all dynamic factors change simultaneously. Section 6.7 summarizes all experiments' findings.

*6.1. The performance of the new CA algorithm in dynamic trustee profiles*

This section presents the results of two experiments, in which we compare the performance of the updated CA algorithm (CA_NEW) with the performance of the old version (CA_OLD), and the performance of FIRE, when the service providers' performance varies over time.

**Experiment 1**. A provider may alter its average performance at maximum 1.0 UG unit with a probability of 10% in every round ($p\mu C = 0.10, M = 1.0$). The results are shown in figure 1. Figure 1a shows that CA_NEW generally outperforms CA_OLD, with the most significant improvement observed in the initial interactions. Figure 1b shows that CA_NEW, except for the first interaction, outperforms FIRE in the initial interactions. However, FIRE adapts as the number of interactions increases and eventually achieves slightly better performance.

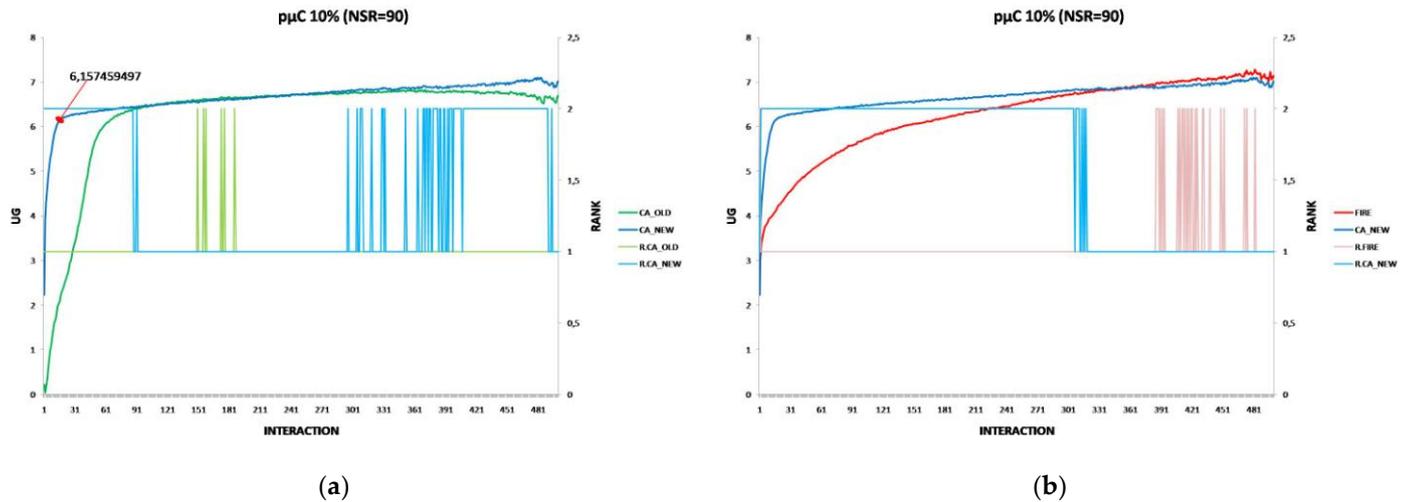

(**a**)          (**b**)

**Figure 1.** (**a**) Performance comparison of CA_NEW and CA_OLD, when providers change performance with 10% probability per round: CA_NEW achieves better performance, especially in early interactions; (**b**) Performance comparison of CA_NEW and FIRE, when providers change performance with 10% probability per round: CA_NEW excels in early interactions (except for the first interaction), while FIRE adapts and slightly outperforms over time.

**Experiment 2**. A provider may switch into a different performance profile with a probability of 2% in every round ($p_{ProfileSwitch} = 0.02$). The results are depicted in figure 2. Figure 2a demonstrates that CA_NEW consistently outperforms CA_OLD in all interactions, with an average difference of 2 UG units. Figure 2b shows that CA_NEW performs better than FIRE in all interactions except the first one.

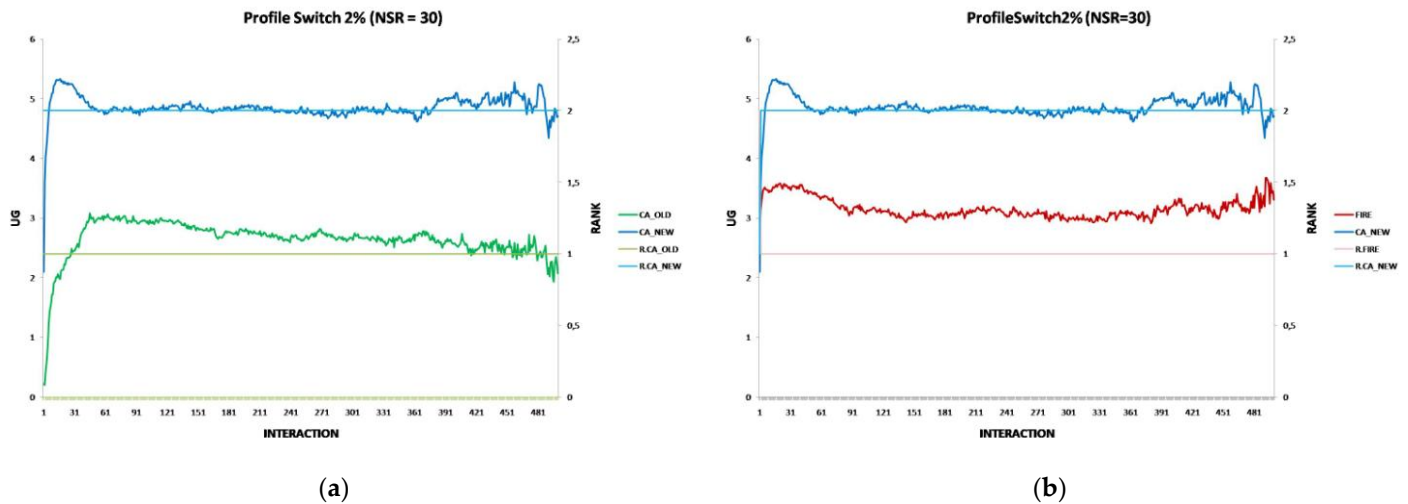

(**a**)          (**b**)

**Figure 2.** (**a**) Performance comparison of CA_NEW and CA_OLD when providers switch performance profiles with 2% probability per round: CA_NEW consistently outperforms CA_OLD by an average of 2 UG units.; (**b**) Performance comparison of CA_NEW and FIRE when providers switch performance profiles with 2% probability per round: CA_NEW outperforms FIRE in all interactions except for the first one.

*6.2. The performance of the new CA algorithm in the static setting*

This subsection presents the results of Experiment 3, which evaluates the performance of the updated CA algorithm (CA_NEW) in a static environment without any dynamic factors. The findings are depicted in Figure 3. Figure 3a demonstrates that CA_NEW performs better than CA_OLD in the initial interactions. Figure 3b shows that CA_NEW surpasses FIRE in all interactions, except for the first one.

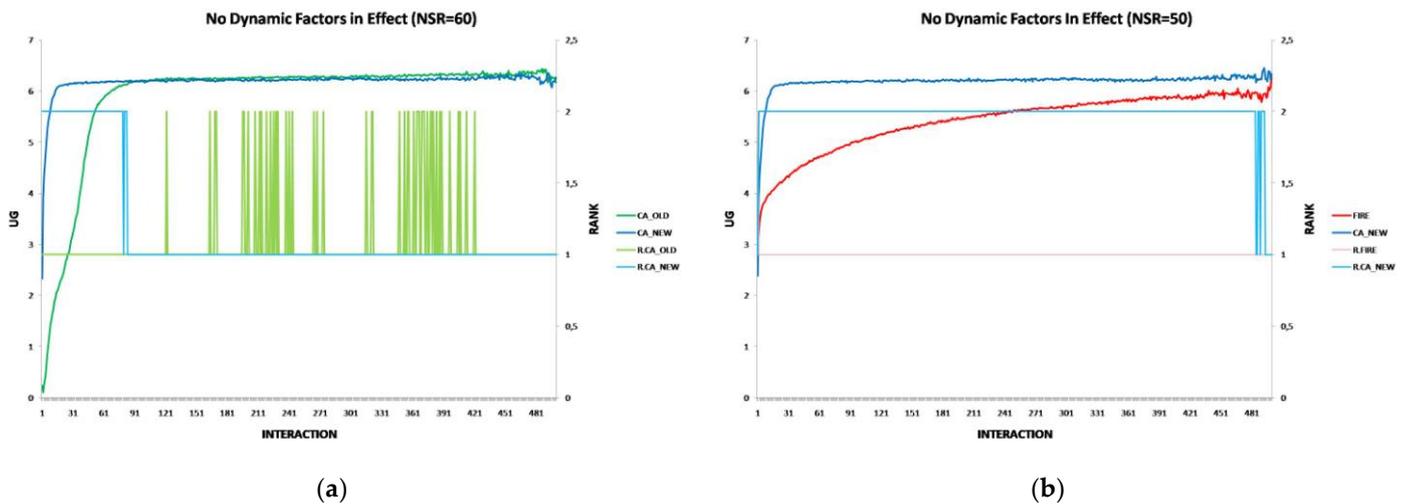

(**a**)          (**b**)

**Figure 3.** (**a**) Performance comparison of CA_NEW and CA_OLD in a static environment: CA_NEW achieves better performance in early interactions; (**b**) Performance comparison of CA_NEW and FIRE in a static environment: CA_NEW outperforms FIRE in all interactions except for the first

*6.3. The performance of the new CA algorithm in provider population changes*

This section evaluates the performance of the updated CA algorithm (CA_NEW) under conditions where the provider population gradually fluctuates up to 30%, through a series of five experiments:

**Experiment 4**. The provider population changes at maximum 2% in every round ($p_{PPC} = 0.02$). The results are shown in figure 4.

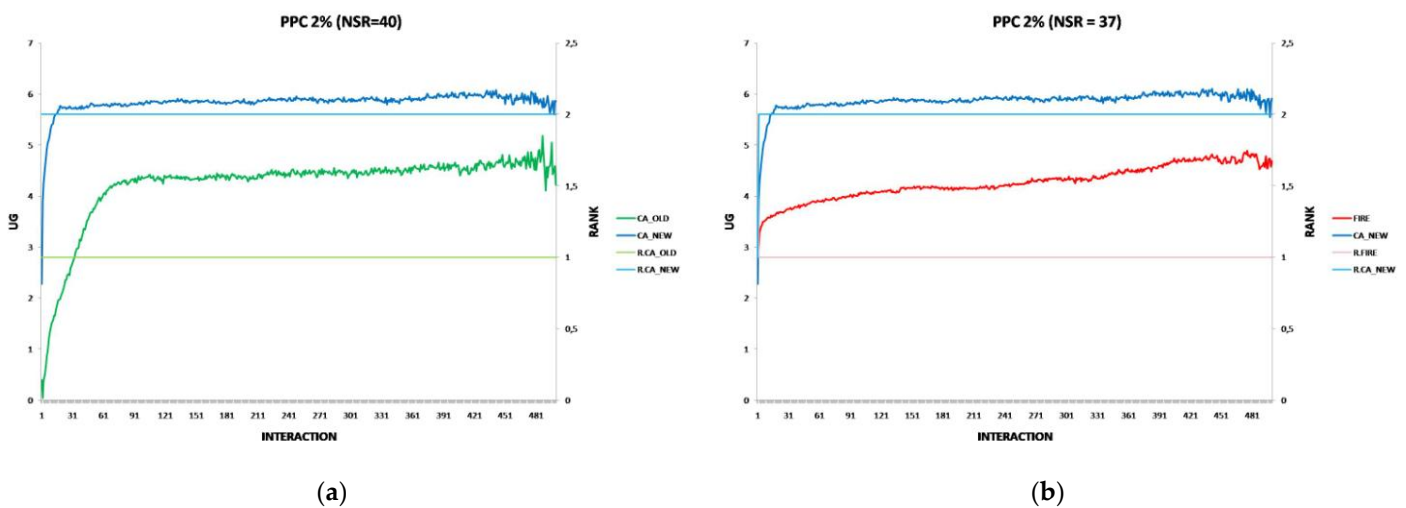

(**a**)          (**b**)

**Figure 4.** (**a**) Performance comparison of CA_NEW and CA_OLD when the provider population changes with 2% probability per round: CA_NEW outperforms in all interactions; (**b**) Performance

comparison of CA_NEW and FIRE when the provider population changes with 2% probability per round.

**Experiment 5**. The provider population changes at maximum 5% in every round ($p_{PPC} = 0.05$). The results are shown in figure 5.

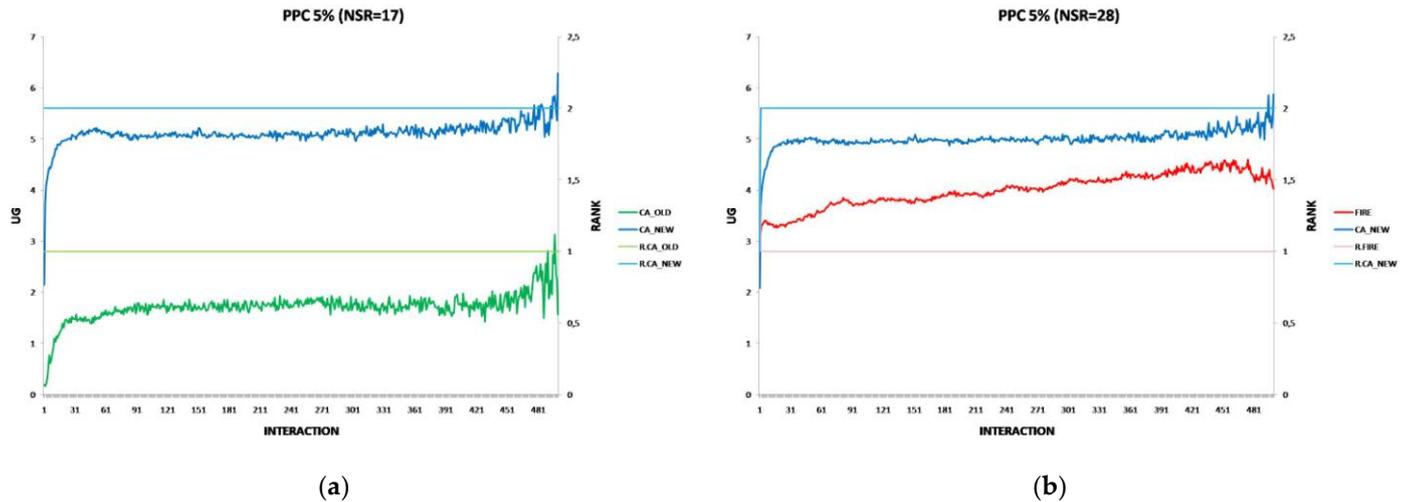

(**a**)                                              (**b**)

**Figure 5.** (**a**) Performance comparison of CA_NEW and CA_OLD when the provider population changes with 5% probability per round: CA_NEW outperforms in all interactions; (**b**) Performance comparison of CA_NEW and FIRE when the provider population changes with 5% probability per round.

**Experiment 6**. The provider population changes at maximum 10% in every round ($p_{PPC} = 0.10$). The results are shown in figure 6.

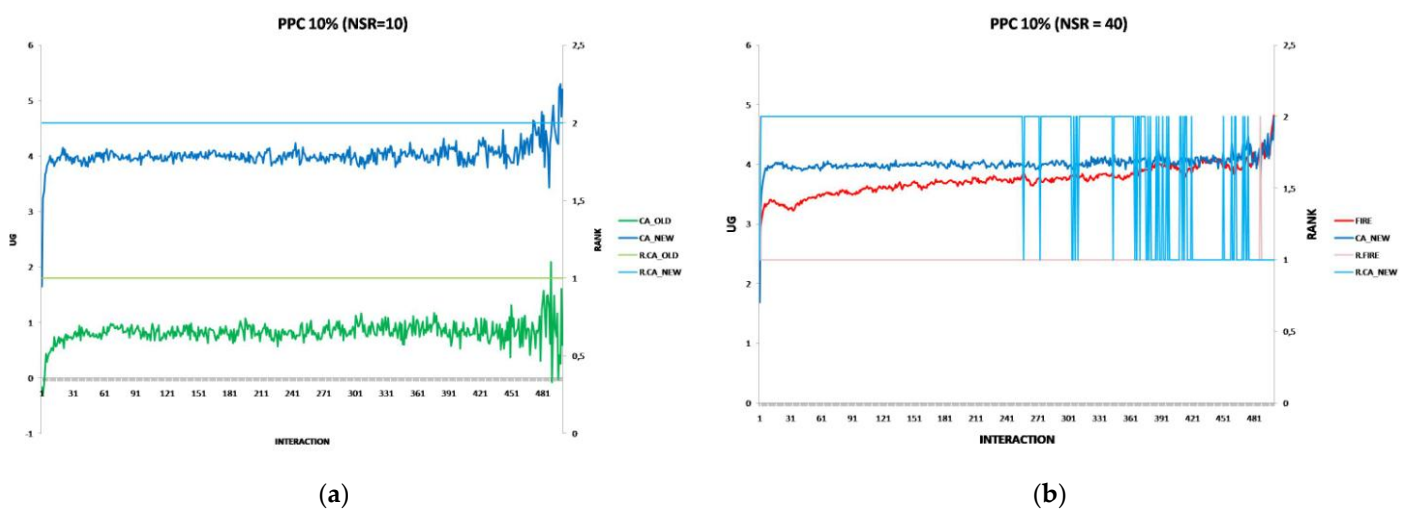

(**a**)                                              (**b**)

**Figure 6.** (**a**) Performance comparison of CA_NEW and CA_OLD when the provider population changes with 10% probability per round: CA_NEW outperforms in all interactions; (**b**) Performance comparison of CA_NEW and FIRE when the provider population changes with 10% probability per round.

**Experiment 7**. The provider population changes at maximum 20% in every round ($p_{PPC} = 0.20$). The results are shown in figure 7.

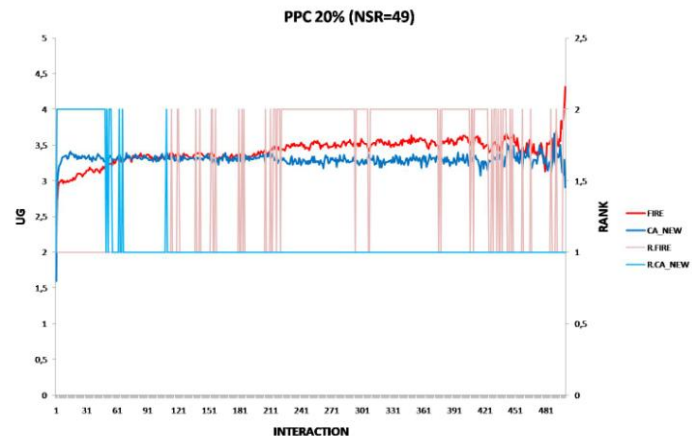

**Figure 7.** Performance comparison of CA_NEW and FIRE when the provider population changes with 20% probability per round.

**Experiment 8**. The provider population changes at maximum 30% in every round ($p_{PPC} = 0.30$). The results are shown in figure 8.

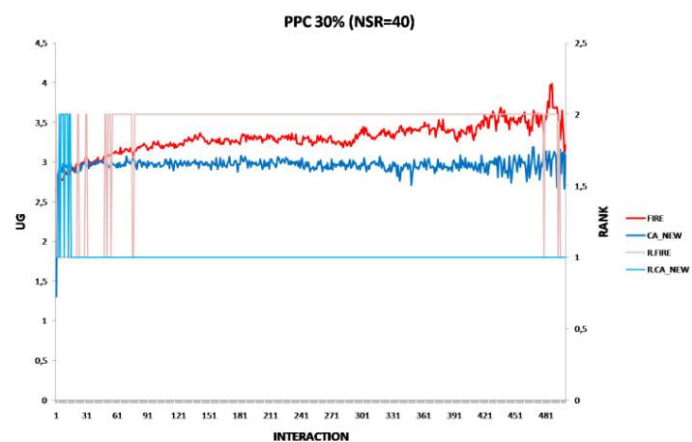

**Figure 8.** Performance comparison of CA_NEW and FIRE when the provider population changes with 30% probability per round.

In Experiments 4-6, we compare the performance of CA_NEW to both CA_OLD and FIRE. Figures 4a, 5a and 6a show that CA_NEW significantly outperforms CA_OLD, achieving higher UG across all interactions. Figures 4b, 5b and 6b reveal that as the provider population change rate rises from 2% to 10%, CA_NEW maintains better performance than FIRE.

Figure 9a and 9b demonstrate that CA_NEW is more resilient than CA_OLD to changes in provider population. Specifically, when the provider population change rate increases from 2% to 10%, CA_OLD's performance drops by 4 UG units (figure 9a), whereas CA_NEW's performance drops only 2 UG units (figure 9b).

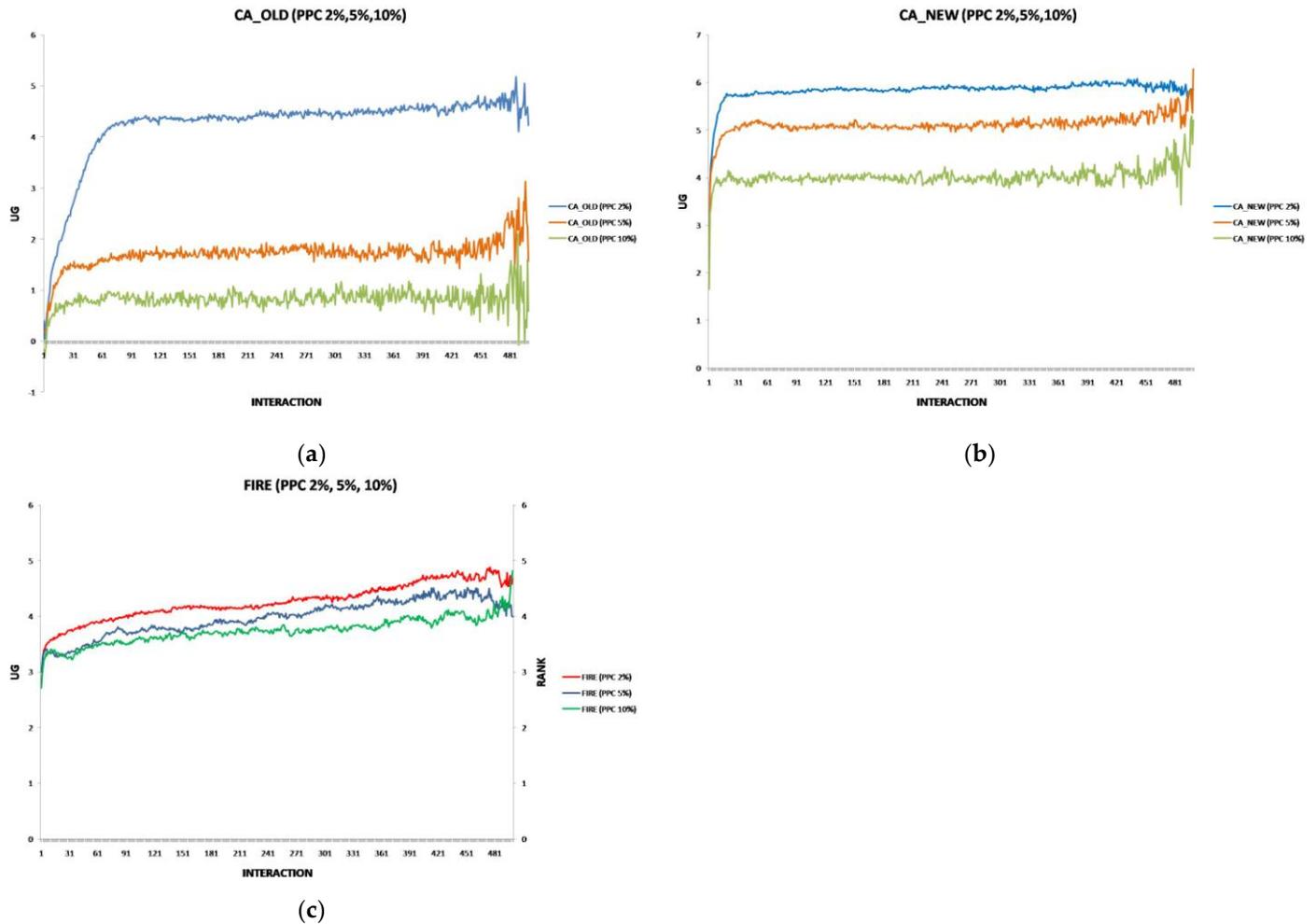

**Figure 9.** (**a**) Performance comparison of CA_OLD when provider population changes with probability 2%, 5% and 10%: Performance drops by 4 UG units; (**b**) Performance comparison of CA_NEW when provider population changes with probability 2%, 5%, 10%: Performance drops by 2 UG units; (**c**) Performance comparison of FIRE when provider population changes with probability 2%, 5%, 10%: FIRE is more resilient than CA to this environmental change.

Figures 9b and 9c indicate that CA_NEW's performance drops more sharply than FIRE'S, suggesting that FIRE is more resilient to this environmental change. This trend raises the expectation that FIRE will eventually surpass CA_NEW at a higher provider population change rate. To verify this hypothesis, we conducted experiments 7 and 8. The results, shown in figures 7 and 8, confirm that when the provider population change rate reaches 30%, FIRE outperforms CA_NEW.

*6.4. The performance of the new CA algorithm in consumer population changes*

This section evaluates the performance of the updated CA algorithm (CA_NEW) in comparison to both CA_OLD and FIRE under conditions where the consumer population gradually fluctuates by up to 10%, through a series of three experiments.

**Experiment 9**. The consumer population changes at maximum 2% in every round ($p_{CPC} = 0.02$). The results are shown in figure 10.

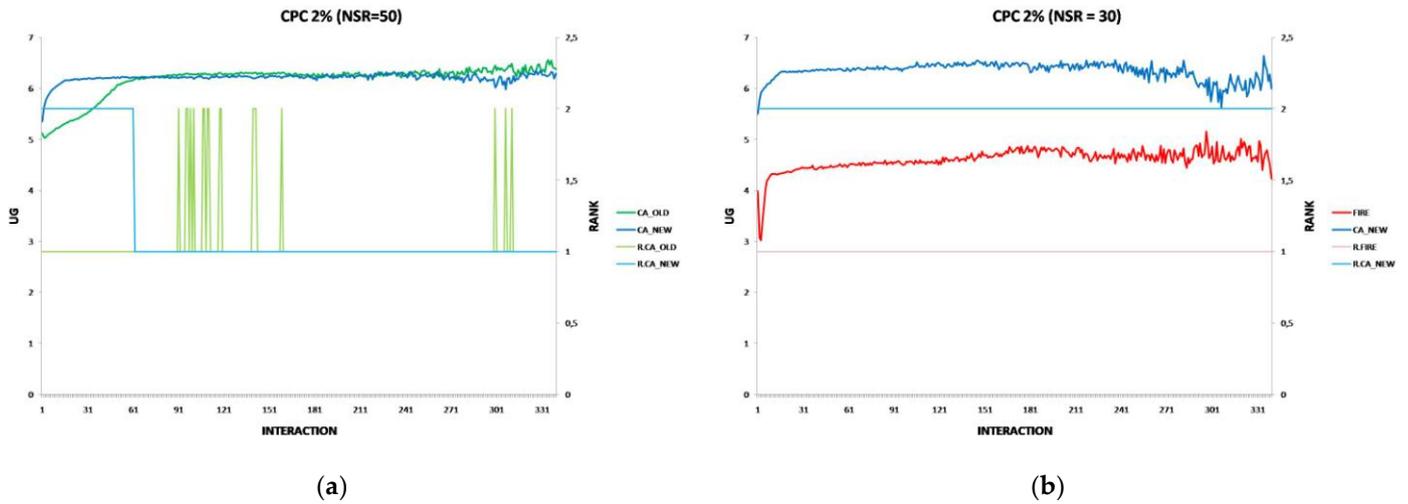

**Figure 10.** (**a**) Performance comparison of CA_NEW and CA_OLD when consumer population changes with probability 2%; (**b**) Performance comparison of CA_NEW and FIRE when consumer population changes with probability 2%.

**Experiment 10**. The consumer population changes at maximum 5% in every round ($p_{CPC} = 0.05$). The results are shown in figure 11.

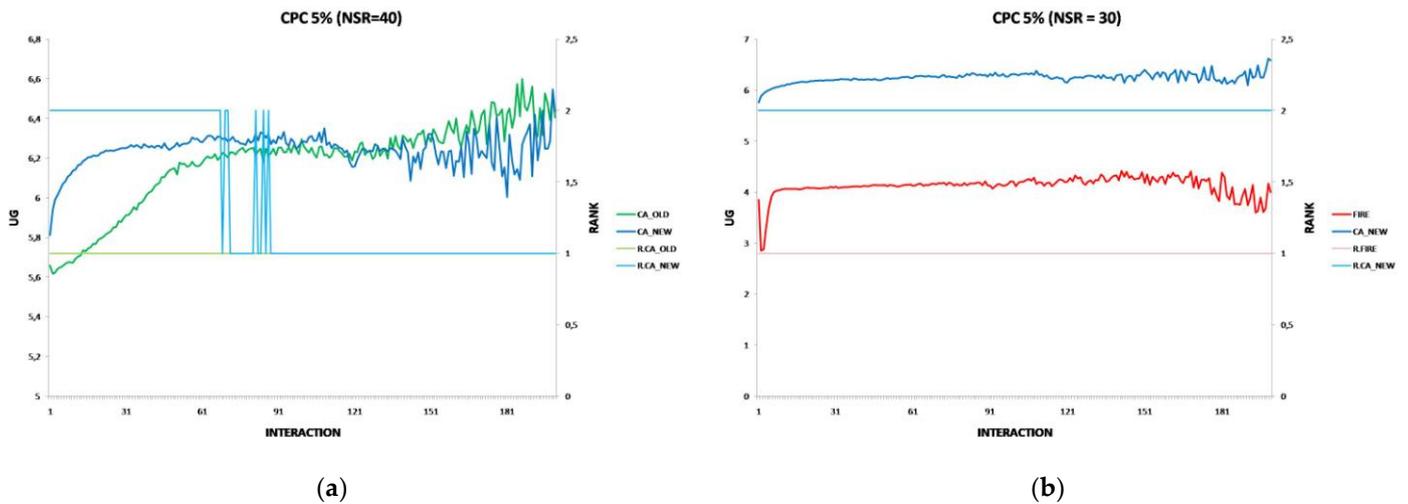

**Figure 11.** (**a**) Performance comparison of CA_NEW and CA_OLD when consumer population changes with probability 5%; (**b**) Performance comparison of CA_NEW and FIRE when consumer population changes with probability 5%.

**Experiment 11**. The consumer population changes at maximum 10% in every round ($p_{CPC} = 0.10$). The results are shown in figure 12.

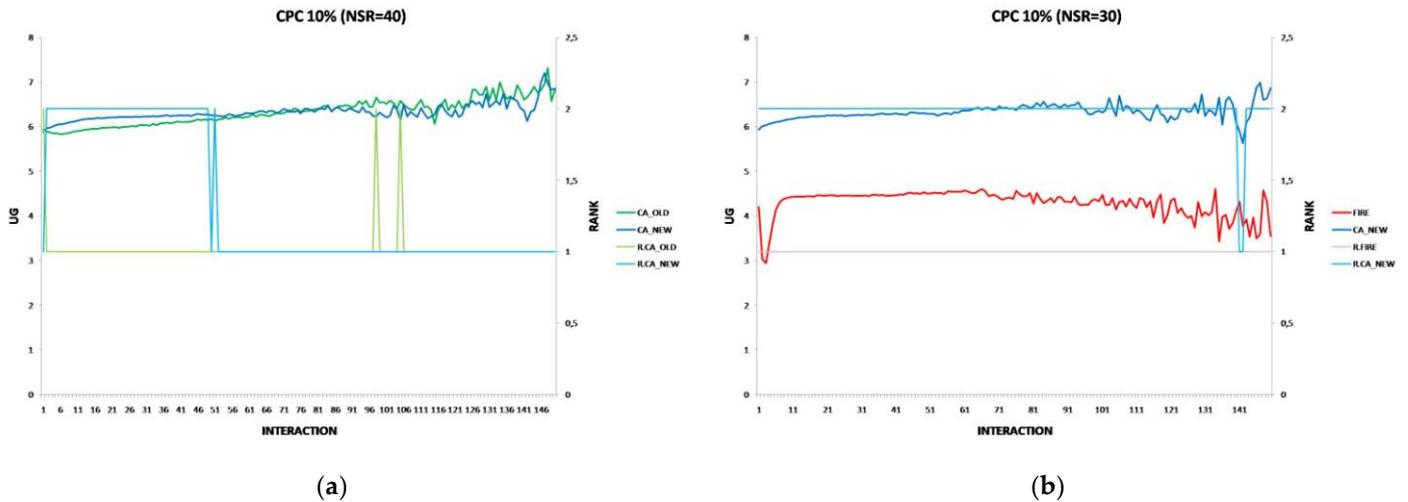

(**a**)　　　　　　　　　　　　　　　　　(**b**)

**Figure 12.** (**a**) Performance comparison of CA_NEW and CA_OLD when consumer population changes with probability 10%; (**b**) Performance comparison of CA_NEW and FIRE when consumer population changes with probability 10%.

Figures 9a, 10a, and 11a illustrate that CA_NEW outperforms CA_OLD, generally achieving higher UG in the initial interactions. Figures 9b, 10b, and 11b demonstrate that CA_NEW consistently performs better than FIRE throughout all interactions in this environmental change.

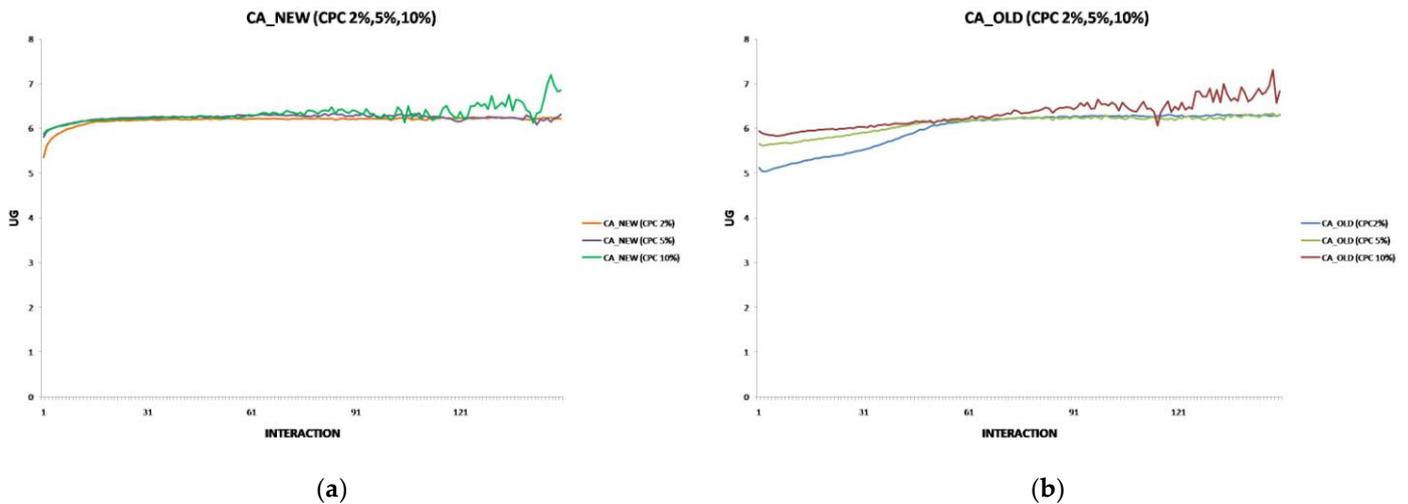

(**a**)　　　　　　　　　　　　　　　　　(**b**)

**Figure 13.** (**a**) Performance comparison of CA_NEW in consumer population changes 2%, 5%, and 10%; (**b**) Performance comparison of CA_OLD in consumer population changes 2%, 5%, and 10%.

Figures 13a and 13b show that in the first interactions both CA_OLD and CA_NEW improve their performance as the consumer population change rate increases from 2% to 10%. Nevertheless, figure 12a reveals that when CPC = 10% CA_OLD slightly surpasses CA_NEW in the first interaction. A possible explanation is that when CPC = 10%, the average UG of the first interaction is influenced by a larger number of newcomer agents who have their first interaction in later simulation rounds. During these rounds, service providers, have established more connections and can more accurately evaluate their ability to provide the service using CA_OLD, resulting in higher UG for service consumers. This observation has lead us to hypothesize that CA_OLD, compared to CA_NEW, is a more suitable choice for service providers that have remained in the sys-tem longer and have gained more knowledge about their service-providing capabilities, assuming their capabilities remain unchanged over time.

*6.5. The performance of the new CA algorithm in consumer and provider location changes*

In this subsection, we evaluate the performance of the updated CA algorithm (CA_NEW) in comparison to both CA_OLD and FIRE under conditions where consumers and providers change locations, by conducting the following two experiments.

**Experiment 12**. A consumer may move to a new location on the spherical world at a maximum angular distance of π/20 with a probability of 0.10 in every round ( $p_{CLC} = 0.10, \Delta\Phi = \pi/20$). The results are shown in figure 14.

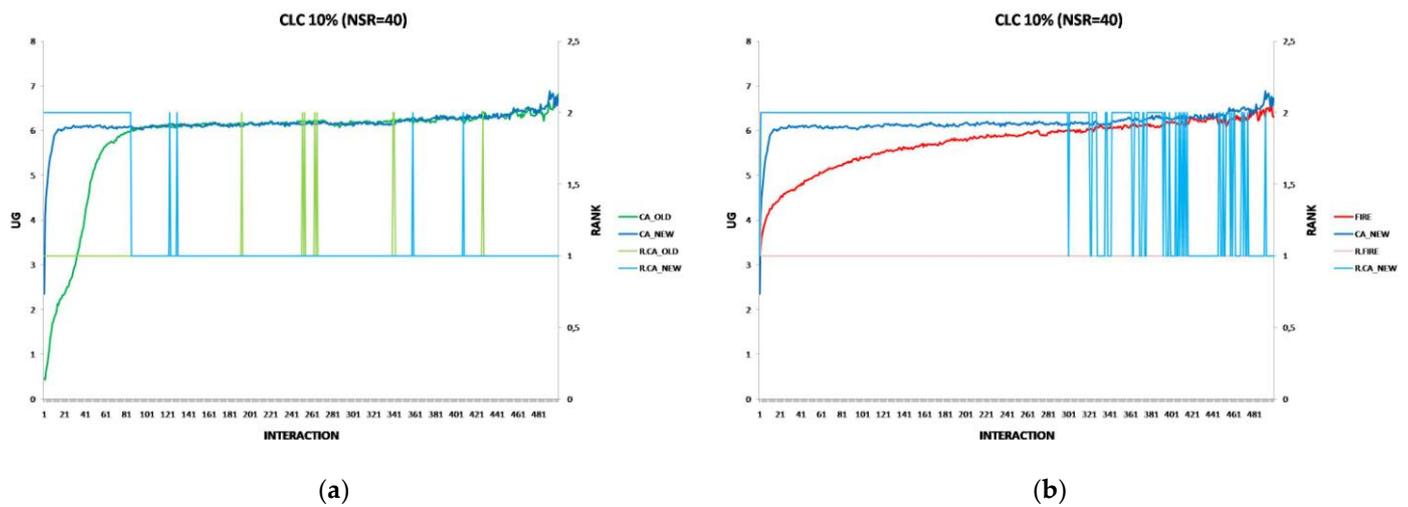

(**a**)           (**b**)

**Figure 14.** (**a**) Performance comparison of CA_NEW and CA_OLD when consumers change locations with a probability of 10% per round; (**b**) Performance comparison of CA_NEW and FIRE when consumers change locations with a probability of 10% per round.

**Experiment 13**. A provider may move to a new location on the spherical world at a maximum angular distance of π/20 with a probability of 0.10 in every round ( $p_{PLC} = 0.10, \Delta\Phi = \pi/20$). The results are shown in figure 15.

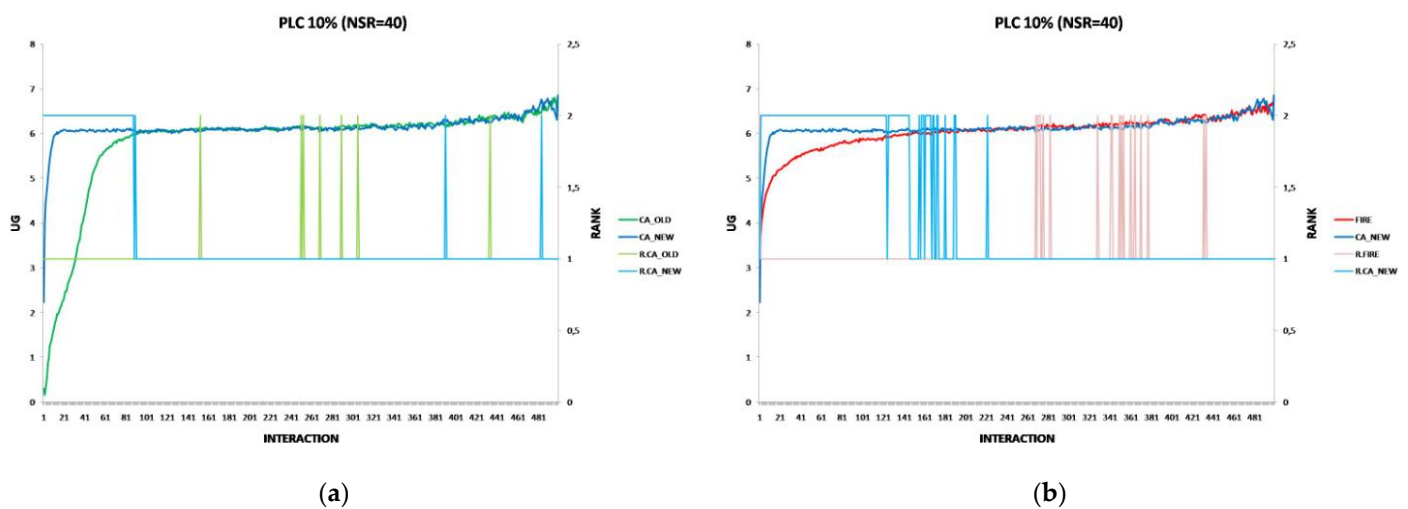

(**a**)           (**b**)

**Figure 15.** (**a**) Performance comparison of CA_NEW and CA_OLD when providers change locations with a probability of 10% per round; (**b**) Performance comparison of CA_NEW and FIRE when providers change locations with a probability of 10% per round.

Figures 14a and 15a indicate that CA_NEW demonstrates improved performance in the initial interactions compared to CA_OLD under both environmental changes. Figures 14b and 15b show that CA_NEW outperforms FIRE in both experiments, except for the first interaction, where FIRE performs better.

*6.6. The performance of the new CA algorithm under the effect of all dynamic factors*

In this subsection, we report the results of the final experiment, which evaluates the performance of CA_NEW under the combined influence of all dynamic factors.

**Experiment 14**. A provider may alter its average performance at maximum 1.0 UG unit with a probability of 10% in every round ($p\mu C = 0.10, M = 1.0$). A provider may switch into a different performance profile with a probability of 2% in every round ($p_{ProfileSwitch} = 0.02$). Provider population changes at maximum 2% in every round ($p_{PPC} = 0.02$). The consumer population changes at maximum 5% in every round ($p_{CPC} = 0.05$). A consumer may move to a new location on the spherical world at a maximum angular distance of $\pi/20$ with a probability of 0.10 in every round ($p_{CLC} = 0.10, \Delta\Phi = \pi/20$). A provider may move to a new location on the spherical world at a maximum angular distance of $\pi/20$ with a probability of 0.10 in every round ($p_{PLC} = 0.10, \Delta\Phi = \pi/20$).

The results, shown in figure 16, demonstrate that CA_NEW consistently outperforms both CA_OLD and FIRE across all interactions.

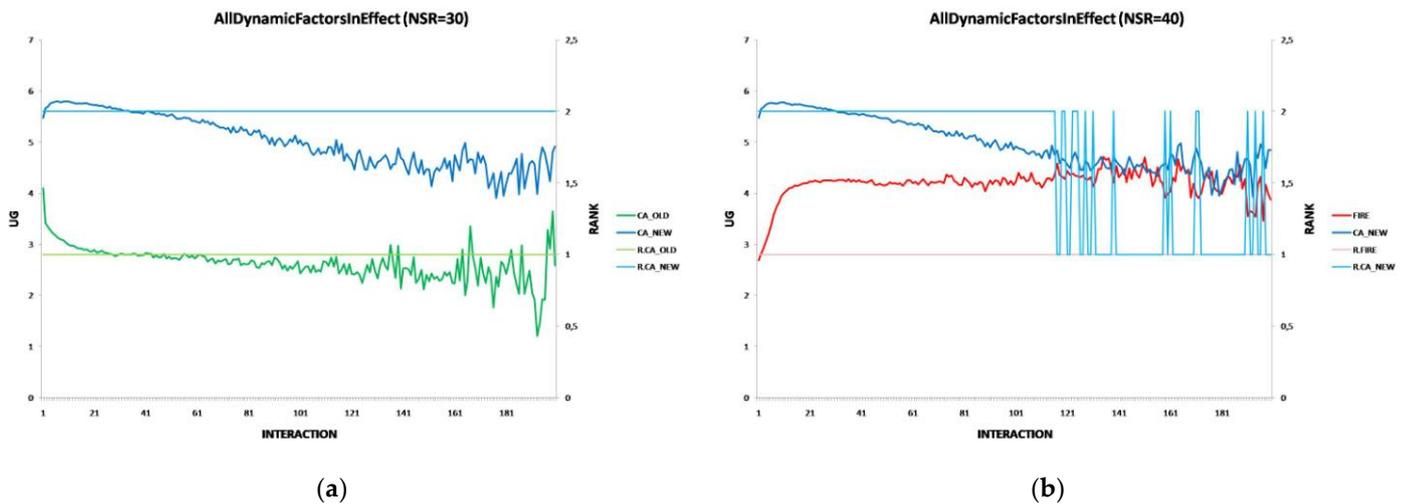

(**a**) (**b**)

**Figure 16.** (**a**) Performance comparison of CA_NEW and CA_OLD when all dynamic factors are in effect: CA_NEW outperforms; (**b**) Performance comparison of CA_NEW and FIRE when all dynamic factors are in effect: CA_NEW outperforms.

*6.7. Overview of the results*

The simulation results reveal that the updated CA algorithm (CA_NEW) demonstrates superior performance over its predecessor (CA_OLD) and the FIRE model under various environmental conditions:

1. Dynamic Trustee Profiles: CA_NEW outperforms CA_OLD across all interactions and shows resilience in handling provider performance fluctuations. While FIRE adapts more quickly in some cases, CA_NEW remains very competitive;
2. Static Environment: CA_NEW surpasses CA_OLD in the initial interactions and consistently outperforms FIRE, except in the first interaction;
3. Provider Population Changes: CA_NEW is more resilient than CA_OLD when provider population fluctuations increase up to 10%, maintaining better performance. However, as changes reach 30%, FIRE eventually outperforms CA_NEW, indicating FIRE's resilience in this environmental change;
4. Consumer Population Changes: CA_NEW generally achieves higher UG than CA_OLD, though in high levels of consumer population changes, CA_OLD performs slightly better in the first interaction, suggesting CA_OLD may be better for old service providers with stable capabilities. CA_NEW consistently outperforms FIRE under consumer population changes;
5. Consumer and Provider Location Changes: CA_NEW shows improved performance over CA_OLD in initial interactions and outperforms FIRE in all interactions, except for the first one;
6. Combined Dynamic Factors: When all dynamic factors are in effect, CA_NEW maintains superior performance over both CA_OLD and FIRE across all interactions, demonstrating its robustness in complex environments.

Overall, CA_NEW is a significant improvement over CA_OLD, with better resilience and adaptability, although there are indications that CA_OLD may be a better choice for old service providers that do not change behavior. While FIRE exhibits some advantages in extreme environmental changes, CA_NEW remains highly competitive across diverse scenarios. Building on previous research [29], where we examined how trustors can identify environmental dynamics and select the optimal trust model (CA or FIRE) to maximize utility, a natural direction for future work is to explore how trustees can recognize environmental changes and assess their own self-awareness to determine whether CA_OLD or CA_NEW would yield the best performance. Adopting this comprehensive RL approach could enhance adaptability and effectiveness across a wide range of scenarios.

## 7. Towards a comprehensive evaluation of the CA model

The authors in [30] define Quality of Trust (QoT) as the quality of trust models and propose several evaluation criteria including Subjectivity, Dynamicity, Context-Awareness, Privacy Preservation, Scalability, Robustness, Overhead, Explainability, and User Acceptance. However, they highlight the absence of a uniform standard. Indeed, in [31], the authors focusing on Edge-based IoT systems, they provide a different set of evaluation criteria (Accuracy, Security, Availability, Reliability, Heterogeneity, Lightweightness, Flexibility, and Scalability) referring to them as trust management requirements, whereas in [32], integrity, computability, and flexibility are identified as key properties of trust models.

This section outlines a comprehensive evaluation of the CA model based on thirteen evaluation criteria for trust models widely discussed in the literature.

**Decentralization.** Centralized cloud-based trust assessment has significant drawbacks, including a single point of failure, scalability issues, and susceptibility to data misuse and manipulation by the companies that own the cloud servers. As a response, decentralized-distributed solutions are becoming the norm in trust management [33]. The CA model follows the decentralized approach by allowing each trustee to compute trust independently, eliminating reliance on a central authority.

**Subjectivity.** Trust is inherently subjective, as different service requesters may have different service requirements in different contexts, affecting their perception for the trustworthiness of the same service provider [32]. Therefore, trust models should take into account the trustor's subjective opinion [30]. CA satisfies the subjectivity requirement, because the service requester defines and broadcasts the task and its requirements and after the task's execution the service requester gives feedback to the service provider rating its performance and the service provider decides if the task was successfully executed, modifying the weight of the relevant connection, based on the feedback received.

**Context Awareness**. Since trust is context-dependent, a service provider may be trustworthy for one task but unreliable for another. The context typically refers to the task type, but it can also refer to an objective or execution environment with specific requirements [30]. CA ensures context awareness by allowing trustees to maintain distinct trust connections for different tasks requested by the same trustor.

**Dynamicity**. Trust is inherently dynamic, continuously evolving in response to events, environmental changes and resource fluctuations. Recent trust information is much more important than historical one [30]. CA model satisfies dynamicity requirement, since trust update is event driven, taking place after each task execution, making it well-suited for dynamic environments, where agents may frequently join, leave, or alter their behavior, as evidenced by the experiments in this study.

**Availability**. A trust model should ensure that all entities and services remain updated [34] and fully available, even in the face of attacks or resource constraints [31]. In the CA approach, the service requester does not assign a task to a specific service provider but instead broadcasts the task to all nearby service providers. This ensures that even if a particular service provider is unavailable due to resource limitations (e.g., battery depletion), another service provider that receives the request can still complete the task. Consequently, our approach effectively meets the availability requirement.

**Integrity and Transparency**. All transaction data and interaction outcomes should be fully recorded, efficiently stored and easily retrieved [32]. Trust values should be accessible to all authorized network nodes whenever they need to assess the trustworthiness of any device in the system [33]. In the CA approach, trust information is stored locally by the trustee in the form of connections (trust relationships). This ensures that trust information remains available even when entities cross application boundaries or transition between systems. Additionally, any trustee can readily access its own trust information to evaluate its reliability for a given task thereby fulfilling the integrity and transparency requirement.

**Scalability**. Since scalability is linked to processing load and time, a trust model must efficiently manage large-scale networks while maintaining stable performance, regardless of network size, and function properly when devices are added or removed [30]. Large-scale networks require increased communication and higher storage capacity, meaning that trust models must adapt to the growing number of nodes and interactions [25, 34, 35]. However, many existing trust management algorithms struggle to scale effectively in massively distributed systems [24]. The CA model is designed for highly dynamic environments, ensuring strong performance despite continuous population changes. Previous research [7] has experimentally demonstrated CA's resilience to fluctuations in consumer populations. In this study, we present simulation experiments showing that the updated CA algorithm has significantly improved resilience, even when provider populations change. Additionally, since agents in the CA approach do not exchange trust information, the model avoids scalability issues related to agent communication. However, we have yet to evaluate how CA scales with an increasing number of nodes.

**Overhead**. A trust model should be simple and lightweight, as calculating trust scores may be impractical for IoT devices with limited computing power [25]. It is essential to ensure that a trust model does not excessively consume a device's resources [23]. Both computational and storage overhead must be considered [30], as devices typically have constrained processing and storage capabilities and must prioritize their primary tasks over trust evaluation. Additionally, excessive computational overhead can hinder real-time trust assessments, negatively impacting time-sensitive applications. A trust model's efficiency can be analyzed using big O notation for time and space complexity. The CA model is considered simple and lightweight since its algorithm avoids complex mathematical computations. However, a detailed analysis of its time and space complexity in big O notation remains to be conducted. Since all trust information is stored locally by the trustees, further research is needed to evaluate the CA model's storage overhead and explore potential optimizations to minimize it.

**Accuracy**. A trust model must effectively identify and prevent malicious entities by ensuring precise classification [25]. It should achieve a high level of accuracy, meaning the computed trust value closely reflects the true value (ground truth) [31]. Several studies [1,10,36,37] assess proposed trust models using metrics such as Precision, Recall, F1-score, Accuracy, False Positive Rate (FPR), True Positive Rate (TPR). However, the accuracy of the CA model has not yet been evaluated using these metrics.

**Robustness**. Trust models must withstand both anomalous behavior caused by sensor malfunctions [35] and trust-related attacks from insider attackers who deliberately act maliciously for personal gain or to disrupt system performance [25,38]. In heterogeneous networks, constant device connectivity and weak interoperability between network domains create numerous opportunities for malicious activities [30]. Traditional cryptographic and authentication methods are insufficient against insider attacks, where attackers possess valid cryptographic keys [35]. Existing trust models can only mitigate certain types of attacks, but none can fully defend against all threats [39]. Therefore, developing algorithms capable of detecting a broad range of malicious activity patterns is crucial [25]. In the following section, we examine common trust-related attacks from the literature and evaluate the CA model's effectiveness in countering them.

**Privacy Preservation**. Ensuring a high level of privacy through data encryption is essential during trust assessment to prevent sensitive information leaks [33]. A trust model must safeguard both Identity Privacy (IP) and Data Privacy (DP) [30]. Specifically, feedback and interaction data must remain protected throughout all stages of data management, including collection, transmission, storage, fusion, and analysis. Additionally, identity details such as names and addresses should be shielded from unauthorized access, as linking trust information to real identities can lead to serious risks, such as Distributed Denial of Service Attack (DDoS) [34]. In the CA approach, while entities generally do not exchange trust information (e.g., recommendations), interaction feedback is transmitted from the trustor to the trustee, allowing the trustee to assess and locally store its own trustworthiness in the form of connections. Therefore, protecting the trustor's identity and ensuring data privacy during transmission, storage, and analysis are critical considerations in the CA approach. Exploring the integration of blockchain technology to enhance privacy protection could be a valuable future research direction.

**Explainability**. Trust models should be capable of providing clear and understandable explanations for their results. The ability to analyze and justify decisions is essential for enhancing user trust, compliance, and acceptance [40]. Additionally, it is important to clarify the processing logic and how trust metrics influence trust evaluations [30]. In this study, we have made two key contributions to improve the explainability of our model. First, we conducted a semi-formal analysis to identify potential modifications to the CA algorithm's processing logic that could enhance its performance. Then, we carried out a series of simulation experiments to demonstrate the effectiveness of these improvements.

**User Acceptance**. The acceptance of a trust model depends on factors such as Quality of Service (QoS), Quality of Experience (QoE), and individual user preferences [30]. It can be assessed by gathering user feedback through questionnaires. However, user acceptance of the CA model has not yet been evaluated.

Overall, the CA model satisfies Decentralization, Subjectivity, Integrity and Transparency, but several aspects require further enhancement and evaluation. Future research should focus on scalability, as the model's performance in large-scale environments with a high number of nodes remains untested. Additionally, while trust data is stored locally, its impact on system overhead and storage efficiency needs further investigation. Accuracy assessment using standard metrics like Precision, Recall, and F1-score is also necessary to validate its reliability. In terms of robustness, although the model resists false recommendations, its resilience against insider attacks requires deeper analysis. Privacy preservation remains an open challenge, particularly in safeguarding trustor identity and feedback transmission, which could benefit from encryption or blockchain-based solutions. Finally, user acceptance has yet to be assessed, making it essential to evaluate the model's adoption based on Quality of Service (QoS) and Quality of Experience (QoE). Addressing these challenges will strengthen the CA model's effectiveness and applicability in dynamic environments. To the best of our knowledge, no other models have undergone such a comprehensive evaluation.

## 8. Trust-related Attacks

In this section, we examine the CA model's resilience against the most common trust-related attacks identified in various research studies [2,16,21,23,24,30,33-35,39,41,42]. A malevolent node is typically defined as a socially uncooperative entity that consistently seeks to disrupt system operations [35]. Its primary objective is to provide low-quality services to conserve its own resources while still benefiting from the services offered by other nodes in the system. Malicious nodes can employ various trust-related attack strategies, each designed to evade detection through different deceptive tactics [39].

**Malicious with Everyone (ME).** In this attack, a node consistently provides low-quality services or misleading recommendations, regardless of the requester. This is one of the most fundamental types of attacks. To counter the ME attack, the CA approach could incorporate a contract-theoretic incentive mechanism. This mechanism would reward honest service providers with utility gains while penalizing dishonest service providers with utility losses, similar to the approach in [19], by awarding or deducting credit coins.

**Bad-mouthing Attack (BMA)**. In this attack, one or more malicious nodes deliberately provide bad recommendations to damage the reputation of a well-behaved node, reducing its chances of being selected as service provider. However, in the CA approach, agents do not exchange trust information through recommendations. Since nodes cannot act as recommenders, our approach is inherently immune to BMA. Most studies assume that service requesters are honest, but as noted in [17], service requesters can also be "ill-intended" or "dishonest" deliberately giving low ratings to a service provider despite receiving good service. Additionally, some service requesters may be "amateur" and incapable of accurately assessing service quality. This represents a specific type of bad-mouthing attack, against which our approach is also resilient. In the CA model, service providers have the autonomy to accept or reject service requests, allowing them to maximize their profits while avoiding dishonest or amateur service requesters. If a service requester unfairly assigns a low rating to a high-quality service, the service provider will respond by decreasing the weight of its connection with that service requester, reducing the likelihood of future interactions. This serves as a built-in mechanism to penalize dishonest service requesters.

**Ballot Stuffing (BSA) or Good-mouthing Attack**. This attack occurs when one or more malicious recommenders (a collusion attack) falsely provide positive feedback for a low-quality service provider to boost its reputation and increase its chances of being selected, ultimately disadvantaging legitimate, high-quality service providers. Since the CA approach does not involve nodes exchanging recommendations, it is inherently resistant to BSA carried out by recommender nodes, which is common in other trust models. However, a specific variation of this attack can occur when an ill-intended or amateur service requester assigns a service provider a higher rating than it actually deserves [17]. In this scenario, while the service provider benefits from an inflated rating, it also suffers by misjudging its actual service capability. Since the CA algorithm requires the service provider to compute the average weight of its existing connections to assess its performance, an inaccurate self-evaluation could lead to financial losses when dealing with honest service requesters. Thus, service providers have a strong incentive to identify and avoid dishonest or inexperienced service requesters. To mitigate this risk, we could implement a mechanism similar to the Tlocal and Tglobal value comparison from [17] or the Internal Similarity concept from [21], where the service provider would evaluate the consistency of a given rating by comparing it to the average or median weights of connections with other service providers for the same service. Alternatively, a contract-theoretic incentive mechanism, as in [18], could deter dishonest service requesters by requiring them to pay a disproportionately high amount for low-quality services received. Additionally, integrating credit quotas (coins) as proposed in [19] could further discourage good-mouthing attacks. The choice of mitigation strategy would depend on the specific application environment in which the CA approach is implemented.

**Self-promoting Attack**. In this attack, a malicious node falsely provides positive recommendations about itself to increase its chances of being chosen as service provider. Once selected, it then delivers poor services or exploits the network. This type of attack is especially effective against nodes that have not previously interacted with the attacker. In the CA approach, service providers do not provide recommendations about themselves, and service requesters do not select service providers. As a result, traditional self-promotion tactics are ineffective. However, a dishonest service provider could still choose to provide a service despite knowing it would harm the service requester. To prevent such behavior, it is crucial to incorporate a penalty mechanism that imposes a financial loss on dishonest service providers. Potential solutions include the fee charge concept from [17], a contract-theoretic mechanism from [18], or an incentive mechanism using credit quotas, such as TrustCoin [19]. The choice of mechanism would depend on the specific application scenario.

**Opportunistic-Service Attack (OSA)**. It occurs when a node manipulates its behavior based on its reputation. When its reputation declines, it offers high-quality services, but once its reputation improves, it provides poor services. This strategy allows the node to sustain a sufficient level of trust to continue being chosen as a service provider. However, in the CA approach, OSA does not enhance a malicious node's chances of being selected as a service provider since trustors do not choose trustees. Consequently, our approach remains resilient against opportunistic-service attacks.

**Sybil Attack (SA)**. This attack occurs when a malicious node generates multiple fake identities to manipulate the reputation of a target node unfairly by providing various ratings. In the CA approach, a malicious service requester could use this tactic to hinder a target service provider's ability to accurately assess its service capability, deplete its resources and reduce its chances of being selected as a service provider. To counter this attack, we can apply the same mechanisms proposed for addressing the good-mouthing attack. Specifically, an approach based on the Internal Similarity concept [21] or a contract-theoretic mechanism based on incentives [18] could either deter attackers from carrying out their attacks or assist the service provider in identifying and isolating malicious service requesters.

**White-washing Attack (WA),** also known as a Re-entry Attack, occurs when a malicious node abandons the system after its reputation drops below a certain threshold and then re-enters with a new identity to erase its negative history and reset its reputation to a default value. In the CA approach, this attack is ineffective because changing a service provider's identity does not impact its likelihood of being selected as a service provider, as trustors do not choose trustees.

**On-Off Attack (OOA)**, also called a Random Attack, is one of the most challenging attacks to detect. In this attack, a malicious node alternates its behavior between good (ON) and bad (OFF) in an unpredictable manner, restoring trust just before launching another attack. During the ON phase, the node builds a positive reputation, which it later exploits for malicious activities. The CA approach is resilient against OOA since trustors do not select trustees, meaning a high reputation does not increase a node's chances of being chosen as a service provider.

**Discrimination or Selective misbehavior Attack**. This kind of attack occurs when an attacker selectively manipulates its behavior by providing legitimate services for certain network tasks while acting maliciously against others. To mitigate this attack in the CA approach, a penalty mechanism should be implemented to impose financial consequences on dishonest service providers. Depending on the application scenario, this could involve the "fee charge concept" [17], a contract-theoretic mechanism [18], or an incentive mechanism using credit quotas like in TrustCoin [19]. One form of this attack is selfish behavior, where service providers prioritize easier tasks that require less effort. Rational agents would evaluate each task based on expected utility – calculating the probability of success multiplied by the reward – and choose accordingly. To counter this in the CA approach, it is crucial to implement incentives that encourage capable agents to take on more challenging tasks. For example, a contract-theoretic mechanism can ensure that highly skilled agents prefer difficult tasks with higher expected utility.

**Denial of Service (DoS) and Distributed Denial of Service (DDoS) Attacks** aim to disrupt a service by overwhelming it with excessive traffic or resource-intensive requests. While DoS attack originates from a single attacker, DDoS attacks involve multiple compromised devices working together to target a single system, making them more difficult to counter. In the CA approach, these attacks remain a potential threat. Therefore, future research should focus on developing effective mechanisms to mitigate DoS and DDoS attacks.

**Storage Attack** occurs when a malicious node manipulates stored feedback data by deleting, modifying, or injecting fake information. In the CA approach, an attacker may attempt to alter or remove the locally stored connections of a trustee. To counter this threat, blockchain technology can be utilized as a safeguard to ensure data integrity and prevent tampering.

## 9. Conclusions and Future Work

Various approaches have been developed to assess trust and reputation in real-world MAS, such as peer-to-peer (P2P) networks, online marketplaces, pervasive computing, Smart Grid, and the Internet of Things (IoT). However, existing trust models face significant challenges, including agent mobility, dynamic behavioral changes, continuous entry and exit of agents and the cold start problem.

To address these issues, we introduced the Create Assemblies (CA) model, inspired by synaptic plasticity in the human brain, where trustees (service providers) can evaluate their own capabilities and locally store trust information, allowing for improved agent mobility handling, reduced communication overhead, resilience to disinformation and enhanced privacy.

Previous work [7] comparing CA with FIRE, a well-known trust model, revealed that CA adapts well to consumer population fluctuations, but was less resilient to provider population changes and continuous performance shifts. This work built on these findings, and used a semi-formal analysis to identify performance pitfalls, which were then addressed by allowing service providers to self-assess if their performance falls below a certain threshold, thereby ensuring faster reaction and better adaptability in dynamic environments.

Simulation results confirm that CA_NEW outperforms the original CA_OLD, with much improved resilience and adaptability, although CA_OLD may still be preferable in scenarios of consumer population change, where long-standing service providers maintain stable performance. While FIRE has certain advantages in extreme environmental changes, CA_NEW remains highly competitive across a wide and diverse variety of environmental conditions. Building on prior research [29], where we explored how trustors can detect environmental dynamics and select the optimal trust model (CA or FIRE) to maximize utility, we propose an obvious direction for future work: exploring how trustees can detect environments dynamics and their self-awareness level to choose between CA_OLD and CA_NEW for optimal performance.

This paper also analyzed CA with respect to established evaluation criteria for trust models and discussed its resilience to most well-known trust-related attacks, proposing countermeasures for dishonest behaviors. While the CA model meets key criteria such as Decentralization, Subjectivity, Context Awareness, Dynamicity, Availability, Integrity and Transparency, it requires further research and improvements. The model's holistic evaluation not only highlights its strengths but also demonstrates a commitment to continuous refinement, positioning it as a highly promising foundation for future trust management solutions.

## Appendix A

*Appendix A.1*

**Proof of Proposition 1**. It suffices to show by mathematical induction that the *n*-th connection for $task_1$ will be initialized to the value of 0.45.

Base Case. (n=2). We know that $BP_1$ has initialized its second connection for $co_2$ to the value 0.45.

Induction hypothesis: Assume that proposition 1 is correct for the k-th connection that $BP_1$ will create for $task_1$.

Induction step: We will show that proposition1 holds for the (k+1)th connection that $BP_1$ will create for $task_1$.

$BP_1$ will calculate the average of its $k$ connections for $task_1$ as follows:

$$average = \frac{w_{co_1} + w_{co_2} + \cdots + w_{co_{k-1}} + w_{co_k}}{k}, \quad (3)$$

where $w_{co_n}$ denotes the weight of the n-th connection of $BP_1$ for $task_1$.

From the induction hypothesis we have that:

$$w_{co_k} = 0.45, \qquad (4)$$

The average of the weights of the k-1 connections is equal to 0.45

$$\frac{w_{co_1} + w_{co_2} + \cdots + w_{co_{k-1}}}{k-1} = 0.45 \Rightarrow w_{co_1} + w_{co_2} + \cdots + w_{co_{k-1}} = (k-1) \cdot 0.45. \qquad (5)$$

From (3), (4), (5) we have : $average = \frac{(k-1) \cdot 0.45 + 0.45}{k} = 0.45$ . □